\documentclass[a4paper,11pt]{article}
\pdfoutput=1 

\usepackage{jheppub} 

\usepackage{url}

\usepackage{mathrsfs,amsbsy,amssymb,latexsym,amsfonts,amsmath,bm}
\usepackage{graphicx,xcolor}
\usepackage{mathtools}
\usepackage{verbatim}
\usepackage[T1]{fontenc}
\usepackage{epsf}
\usepackage[utf8]{inputenc}

\newcommand{\tr}{\text{tr} }
\newcommand{\dd}{{\rm d}}

\newcommand{\Ga}{\Gamma}
\newcommand{\la}{\lambda}
\newcommand{\ga}{\gamma}

\newcommand{\cD}{\mathcal D}
\newcommand{\cF}{\mathcal F}
\newcommand{\cG}{\mathcal G}

\newcommand{\cL}{\mathcal L}
\newcommand{\cN}{\mathcal N}
\newcommand{\cO}{\mathcal O}\newcommand{\cP}{\mathcal P}

\newcommand{\cY}{\mathcal Y}

\newcommand{\sfa}{{\mathsf{a}}}

\newcommand{\sfi}{{\mathsf{i}}}

\newcommand{\sfr}{{\mathsf{r}}}
\newcommand{\sfs}{{\mathsf{s}}}
\newcommand{\sfA}{{\mathsf{A}}}

\newcommand{\sfC}{{\mathsf{C}}}

\newcommand{\fC}{{\mathfrak C}}

\newcommand{\chX}{\check{X}}
\newcommand{\chA}{\check{A}}
\newcommand{\chB}{\check{B}}
\newcommand{\chC}{\check{C}}

\newcommand{\hX}{\hat{X}}
\newcommand{\hA}{\hat{A}}
\newcommand{\hB}{\hat{B}}
\newcommand{\hC}{\hat{C}}

\newcommand{\hW}{\hat{W}}
\newcommand{\hnabla}{\hat{\nabla}}
\newcommand{\hsfA}{\hat{\sfA}}
\newcommand{\sfH}{\hat{\sfH}}

\newcommand{\hcD}{\hat{\cD}}
\newcommand{\hsfC}{\hat{\sfC}}

\newcommand{\no}{\nonumber}
\newcommand{\bG}{{\bf{G}}}

\title{\boldmath Superconformal Block from Holographic Geometry.}

\author[a]{Heng-Yu Chen}
\author[a,b]{and Jun-ichi Sakamoto}

\affiliation[a]{Department of Physics, National Taiwan University, Taipei 10617, Taiwan}
\affiliation[b]{Osaka City University Advanced Mathematical Institute (OCAMI), 3-3-138, Sugimoto,
Sumiyoshi-ku, Osaka, 558-8585, Japan}

\emailAdd{heng.yu.chen@phys.ntu.edu.tw}
\emailAdd{sakamoto@ntu.edu.tw}

\abstract{We explicitly construct the holographic dual configuration for the four dimensional $\cN=4$ superconformal block containing half-BPS scalar primary operators by considering its full $AdS_5 \times S^5$ dual geometry. We extend the embedding space formalism and the related Harmonic analysis to general $d$-dimensional sphere $S^d$, and obtain precisely the $R$-symmetry contribution to the half-BPS scalar superconformal blocks, which we refer as ``$R$-symmetry block''. We also observe that the $R$-symmetry quadratic Casimir operator can be mapped to BC$_{2}$ Calogero-Sutherland system Hamiltonian, such that $R$-symmetry block is in terms identified as its bound state wave function.}

\begin{document}

\maketitle
\flushbottom

\section{Introduction and Summary}
\paragraph{}
Among the various developments in studying AdS/CFT correspondence, one of the most common approaches has been reformulating various field theoretical observables into geometrical objects in their dual gravitational theories. This approach is particularly powerful when the class of observables considered are constrained by their underlying symmetries, and often provides more efficient and intuitive ways to compute them. Moreover, such a geometrization of field theoretical observables usually yields interesting alternative representations which lead to new physical insights or non-trivial mathematical connections. 
\paragraph{}
The so-called ``conformal blocks'' or the closely related ``conformal partial waves''\footnote{To be precise, in this work we are considering ``global conformal block'', rather than the much more intricate ``Virasoro conformal blocks'' in two dimensional CFTs.}, which are the universal building ingredients of the conformal correlation functions, precisely belong to this class of observables. They play the role of harmonic functions in conformal field theories, and their explicit form is governed kinematically by the well-known conformal Casimir equation {\cite{Dobrev:1977qv} (see also \cite{Dolan:2011dv}, \cite{Dolan:2003hv})}.
Finding the holographic dual configuration of $d$-dimensional conformal partial wave in the $d+1$ dimensional Anti-de Sitter space (AdS) can be regarded as geometrizing the conformal Casimir equation, indeed it has close relation with the AdS harmonic equation. This task was performed in an elegant paper \cite{Hijano:2015zsa}, and the resultant configuration is now known as ``geodesic Witten diagram''. As its name suggests, the interaction vertices of geodesic Witten diagrams are restricted to move along the AdS geodesics connecting the boundary insertion points of the CFT primary operators. In a somewhat parallel but equally exciting development, it was discovered that conformal Casimir equation can be directly identified with the eigenvalue equation of BC$_{2}$ Calogero-Sutherland quantum integrable system \cite{Isachenkov:2016gim,Isachenkov:2017qgn, Buric:2019rms}, hence the conformal block and the eigenfunction. The BC$_{2}$ Calogero-Sutherland eigenvalue equation can be further mapped to so-called Heckman-Opdam hypergeometric 
equation, whose solution can be constructed in terms of the Harish-Chandra series. Using this chain of relations, the conformal blocks can be constructed as the scattering solutions to the BC$_{2}$ Calogero-Sutherland system and their properties such as monodromy can be systematically studied. We will review some of these details in the later section.
\paragraph{}
It is interesting to ask if we can extend the story we summarize so far about the conformal block/partial wave into full string theoretic constructions, in other words we need to consider the compact manifolds as well as the non-compact AdS spaces.
This also requires us to generalize the conformal symmetries to much richer superconformal symmetries, and consider the holographic dual configurations which reproduce the superconformal blocks/partial waves. In this work, we focus on the prototype of AdS/CFT correspondence, namely the exact correspondence between four dimensional ${\mathcal{N}}=4$ super Yang-Mills theory in the planar limit and Type-IIB superstring theory on $AdS_5 \times S^5$ background. In the nascent days of AdS/CFT, various Witten diagram computations have been performed and matched with the appropriate correlation functions, see e.g. \cite{Lee:1998bxa}, \cite{Arutyunov:1999en} and \cite{DHoker:2002nbb} for review.
Here we would like to fill the gap in the literature by explicitly constructing the holographic dual configuration that reproduce the superconformal blocks involving exclusively half-BPS operators. This is the simplest type of the four dimensional ${\mathcal{N}}=4$ superconformal blocks, whose corresponding superconformal Casimir equation has been constructed in \cite{Nirschl:2004pa,Dolan:2004mu}. For this particularly simple choice, the superconformal Casimir equation can be separated into two parts: one for global conformal symmetries, and the other one for $R$-symmetries.
To geometrize this equation, we first extend the embedding formalism for AdS space \cite{Costa:2014kfa} to general $d$-dimensional sphere, while some of the details are quite similar there remain various interesting subtleties as we will discuss\footnote{We should mention that some simpler bulk computations were also done in an earlier literature \cite{Uruchurtu:2011wh}, here we fully generalize them using embedding formalism and construct the split representation of spherical harmonics.}. 
In particular, we will generalize the split integral representation of AdS harmonic function to the spherical one for arbitrary spin, and demonstrate it satisfies the appropriate equation of motion. 
Using this and combining with the AdS contribution to explicitly construct the holographic dual of the desired superconformal partial waves. Moreover, as an interesting observation, we also notice that the quadratic Casimir operator for the $R$-symmetries can also be mapped to BC$_{2}$ Calogero-Sutherland system Hamiltonian. This times due to the compact $R$-symmetry group, the eigenfunctions correspond to the bound state solutions instead of the scattering ones.
\paragraph{}
This work is organized as follows: In section \ref{Sec:4pt-Review}, we review the the necessary details about the superconformal partial waves/blocks for half-BPS scalar primary operators in four dimensional $\cN=4$ SYM, and their superconformal Casimir equations. This also allows us to fix the notations used throughout this work. In section \ref{Sec:Harmonic}, we systematically construct the holographic dual of the half-BPS superconformal block by extending the embedding space formalism to general $d$-dimensional sphere.
Using the split representation of spherical Harmonic function, we obtain what we called the $R$-symmetry block which is a solution to the $R$-symmetry part of superconformal Casimir equation. In Section \ref{sec:HO}, we discuss how superconformal Casimir equation for half-BPS primary operators can be mapped to two copies of BC$_{2}$ Calogero-Sutherland system. We also review related details about Heckman-Opdam systems and Harish-Chandra hypergeometric functions. We have streamlined our main text, and relegated various computational details in several appendices. 

\section{Four-point functions with half-BPS operators in $\cN=4$ SYM}\label{Sec:4pt-Review}
\paragraph{}
In this section, we will summarize the essential details about four-point correlation functions in four dimensional $\cN=4$ Super Yang-Mills (SYM) with $SU(N)$ gauge group in the so-called planar limit such that $N \to \infty$ and `t Hooft coupling $\lambda = g_{\rm YM}^2 N$ is kept fixed and arbitrary.

\subsection{Four-point functions of half-BPS operators}\label{subsec:setup}

Let us focus on the single-trace half-BPS operators in four dimensional $\cN=4$ $SU(N)$ SYM, which are the gauge invariant operators constructed from the elementary scalar fields $\phi_{\hA}(x)\,(\hA=1,\dots, 6)$ transforming in the adjoint representation of $SU(N)$:
\begin{align}
\cO_{p}(x,T)\equiv T^{\hA_1}\dots T^{\hA_p}\, \tr\left(\phi_{\hA_1}(x)\dots \phi_{\hA_p}(x)\right)\,,\qquad p\geq2\,.
\label{def:halfBPSop}
\end{align}
Here $T^{\hA}\in \mathbb{C}^6$ is an auxiliary complex null vector for performing the tensor contraction and ensuring the traceless condition.
Since the scalar fields $\phi_{\hA}(x)$ transform as the vectors of $SO(6)_R \cong SU(4)_R$\,, the single trace operator (\ref{def:halfBPSop}) describes a rank-$p$ $SO(6)_R$ symmetric traceless tensor.
In terms of the Dynkin labels of $SU(4)_R$\,, (\ref{def:halfBPSop}) belongs to $[0,p,0]$\, representation.
This operator (\ref{def:halfBPSop}) {belongs to a short multiplet \cite{Dobrev:1985qv} of $SU(2,2|4)$ that }is annihilated by half of the Poincar\'e supercharges, so it has the scaling dimension $\Delta=p$ which is protected from any quantum correction.

\paragraph{}

Let $\bG(x_i,T_i)$ denotes four-point correlation function for $\cO_{p_i}(x_i,T_i)$, i.~e.~
\begin{align}
\bG(x_i,T_i)&=\langle\cO_{p_1}(x_1,T_1)\cO_{p_2}(x_2,T_2)\cO_{p_3}(x_3,T_3)\cO_{p_4}(x_4,T_4)\rangle\,.
\label{eq:4pt-1/2BPS}
\end{align}
This is the lowest component of a four-point correlation function of superfields considered in \cite{Bissi:2015qoa}, \cite{Dolan:2004iy}.
If we expand the four-point function (\ref{eq:4pt-1/2BPS}) in the $s$-channel OPE,
the conformal and the $R$-symmetries allow us to fix the form of  $\bG(x_i,T_i)$ into 
\begin{align}
\bG(x_i,T_i)=
\left(\frac{T_{12}}{x^2_{12}}\right)^{\frac{p_1+p_2}{2}}
\left(\frac{T_{34}}{x^2_{34}}\right)^{\frac{p_3+p_4}{2}}
\left(\frac{x^2_{14}T_{24}}{x^2_{24}T_{14}}\right)^{a}
\left(\frac{x^2_{14}T_{13}}{x^2_{13}T_{14}}\right)^{b}\,
\cG^{a,b}(u,v,\sigma,\tau)\,,
\label{eq:4pt-s}
\end{align}
where $x_{ij}\equiv x_i-x_j$ and $T_{ij}\equiv -2T_{i}\cdot T_j$ and we defined\footnote{We can also use the Poincare embedding to express $T_{ij}^2$ as $y_{ij}^2$ \eqref{eq:T-y}.}
\begin{align}
a=-\frac{p_{12}}{2}\,,\qquad b=\frac{p_{34}}{2}\,, \quad p_{ij}=p_i-p_j.
\end{align}
The conformal cross ratios $u$ and $v$ are defined as
\begin{align}
u=\frac{x^2_{12}x^2_{34}}{x^2_{13}x^2_{24}}=z_1\,z_2\,,\qquad
v=\frac{x^2_{23}x^2_{14}}{x^2_{13}x^2_{24}}=(1-z_1)(1-z_2)\,.
\end{align}
Similarly, we can introduce the $R$-symmetry cross ratios $\sigma\,,\tau$ as
\begin{align}
\sigma&=\frac{T_{12}T_{34}}{T_{13}T_{24}}=\alpha_1\,\alpha_2\,,\qquad 
\tau=\frac{T_{23}T_{14}}{T_{13}T_{24}}=(1-\alpha_1)(1-\alpha_2)\,.
\label{def:R-sym-ratio}
\end{align}
Here for the $R$-symmetry cross ratios, we follow the notation of \cite{Bissi:2015qoa}
\footnote{Note that the definitions of $\sigma, \tau$ are slightly different from the conventional one introduced in \cite{Nirschl:2004pa}.
The two different $R$-symmetry cross ratios are related by
\begin{align}
\sigma|_{\text{here}}&=\frac{1}{\sigma}\Bigl|_{\text{Nirschl-Osborn}}\,,\qquad
\tau|_{\text{here}}=\frac{\tau}{\sigma}\Bigl|_{\text{Nirschl-Osborn}}\,.
\end{align}
In $(\alpha_1,\alpha_2)$\,, this relation is translated to
\begin{align}
\alpha_{i}|_{\text{here}}&=\frac{1}{\alpha_{i}}\Bigl|_{\text{Nirschl-Osborn}}\,.
\end{align}
The advantage of the definition is that the conformal block and the $R$-symmetry block can be treated on the equal footing.}.

\medskip

The cross-ratio dependent function $\cG^{a,b}(u,v,\sigma,\tau)$ can be expanded in terms of the lowest component $\cG^{a,b}_{\Delta,J,l,s}(u,v,\sigma,\tau)$ of the superconformal block associated with the superconformal primary operator with scaling dimension $\Delta$\,,  spin $J$ and the $SU(4)$ Cartan numbers $[s,l-s,s]$. 
More explicitly, the expansion of four point correlation function $\cG^{a,b}(u,v,\sigma,\tau)$ into conformal blocks $\cG^{a,b}_{\Delta,J,l,s}(u,v,\sigma,\tau)$ associated with the exchanged operator labeled by $(\Delta, J, l, s)$ has the following form:
\begin{align}
\cG^{a,b}(u,v,\sigma,\tau)=\sum_{\Delta,J,l,s}c(\Delta,J,l,s)\,
\cG^{a,b}_{\Delta,J,l,s}(u,v,\sigma,\tau)\,.
\end{align}
The exchanged operators should be in the irreducible representations of $SU(4)_{R}$-symmetry which appear in both decompositions of $[0,p_1,0] \otimes [0,p_2,0]$ and $[0,p_3,0] \otimes [0,p_4,0]$\,.
From the decomposition rule (\ref{eq:SU(4)-rep-split}) of $SU(4)$\,, the summations of $l$ and $s$ are taken over the values:
\begin{align}
0\leq l \leq \text{min}(p_i)\,,\quad
\text{max}(|p_{12}|, |p_{34}|)+l\leq s \leq \text{min}(p_1+p_2,p_3+p_4)-l\,.
\end{align}
The coefficient $c(\Delta,J,l,s)$ is the product of the three-point function coefficients of two external half-BPS operators and one intermediate superconformal primary operator.

\subsection{Superconformal Casimir equation for the lowest component}\label{subsec:SCE}

The superconformal block is defined as an eigenfunction of the suitable quadratic Casimir equation derived from $PSU(2,2|4)$ superconformal group.
{As explained in {\cite{Nirschl:2004pa,Dolan:2004mu,Bissi:2015qoa}}}, the superconformal Ward identities can then relate the lowest component $\cG^{a,b}_{\Delta,J,l,s}(u,v,\sigma,\tau)$ of the superconformal block with its superconformal descendants.
By combining the superconformal Ward identities and the quadratic superconformal Casimir equation, we obtain the quadratic differential equation for {$\cG^{a,b}_{\Delta,J,l,s}(u,v,\sigma,\tau)$\, {\cite{Nirschl:2004pa,Dolan:2004mu,Bissi:2015qoa}},}
\begin{align}
\cD_{\mathfrak{b}}^{(a,b)} \cG^{a,b}_{\Delta,J,l,s}+\cD_{\mathfrak{f}}\,\cG^{a,b}_{\Delta,J,l,s}=\frac{1}{2}C_{\Delta,J,l,s}\,\cG^{a,b}_{\Delta,J,l,s}\,,\label{eq:superCsimireq}
\end{align}
where $C_{\Delta,J,l,s}$ is the eigenvalue of the quadratic Casimir operator $\fC$ for $PSU(2,2|4)$ given by
\footnote{See equation (23) in \cite{Bissi:2015qoa}, for the $SU(4)$ presentation $[r_1,q,r_2]$\,. }
\begin{align}
C_{\Delta,J,l,s}=\Delta(\Delta+4)+J(J+2)-l(l+4)-s(s+2)\,.
\label{eq:SCas-ev}
\end{align}
{The eigenvalue given above consists of two contributions: the first two terms come from the eigenvalue for $SO(2,4)$ Casimir operator with the shift: $\Delta\to \Delta+4$\,; the remaining two terms come from the eigenvalue of $SO(6)$ Casimir operator.}
\paragraph{}
Let us give explicit expressions of the differential operators $\cD_{\mathfrak{b}}^{(a,b)}$ and $\cD_{\mathfrak{f}}$\,.
For convenience, we will decompose the differential operator $\cD_{\mathfrak{b}}^{(a,b)} $ into the contributions from $SO(2,4)$ and $SO(6)$ subgroups,
\begin{align}
\cD_{\mathfrak{b}}^{(a,b)} &=\check{\cD}_{4}^{(a,b)}-\hat{\cD}_{4}^{(a,b)}\,,
\label{eq:D0-Casimir}
\end{align}
where $\check{\cD}_{d}^{(a,b)}$ and $\hat{\cD}_{d}^{(a,b)}$ are the quadratic Casimir equations for $SO(2,d)$ and $SO(d+2)$ given by:
\begin{align}
\check{\cD}_{d}^{(a,b)}&=\sum_{i=1}^2\left(z_i^2(1-z_i)\frac{\partial^2}{\partial z_i^2}
-(a+b+1)z_i^2\frac{\partial}{\partial z_i}-ab\,z_i\right)\no\\
&\quad+(d-2) \frac{z_1z_2}{z_1-z_2}
\left((1-z_1)\frac{\partial}{\partial z_1}-(1-z_2)\frac{\partial}{\partial z_2}\right)\label{defchD}
\,,\\
\hat{\cD}_{d}^{(a,b)}&=\sum_{i=1}^2\left(\alpha_i^2(1-\alpha_i)\frac{\partial^2}{\partial \alpha_i^2}
+(a+b-1)\alpha_i^2\frac{\partial}{\partial \alpha_i}-ab\,\alpha_i\right)\no\\
&\quad+(d-2) \frac{\alpha_1\alpha_2}{\alpha_1-\alpha_2}
\left((1-\alpha_1)\frac{\partial}{\partial \alpha_1}-(1-\alpha_2)\frac{\partial}{\partial \alpha_2}\right)\,.\label{defhatD}
\end{align}
The differential operator $\cD_{\mathfrak{f}}$ consists of the first order derivatives with respect to $z_{i}$ and $\alpha_{i}$\,, 
\begin{align}
\scalebox{0.9}{$\displaystyle\cD_{\mathfrak{f}}$}&=\scalebox{0.9}{$\displaystyle
\frac{2z_1(z_1-1)(z_1(\alpha_1+\alpha_2)-2\alpha_1\alpha_2)}{(z_1-\alpha_1)(z_1-\alpha_2)}\frac{\partial}{\partial z_1}
+\frac{2z_2(z_2-1)(z_2(\alpha_1+\alpha_2)-2\alpha_1\alpha_2)}{(z_2-\alpha_1)(z_2-\alpha_2)}\frac{\partial}{\partial z_2}
$}\no\\
&\quad\scalebox{0.9}{$\displaystyle 
-\frac{2\alpha_1(\alpha_1-1)(\alpha_1(z_1+z_2)-2z_1z_2)}{(\alpha_1-z_1)(\alpha_1-z_2)}\frac{\partial}{\partial \alpha_1}
-\frac{2\alpha_2(\alpha_2-1)(\alpha_2(z_1+z_2)-2z_1z_2)}{(\alpha_2-z_1)(\alpha_2-z_2)}\frac{\partial}{\partial \alpha_2}\,.
$}
\end{align}
This operator comes from the fermionic part of the quadratic superconformal Casimir equation and is anti-symmetric under the exchange of variables $z_i\leftrightarrow \alpha_i$\,.

\subsection{Superconformal block for the long multiplets}\label{subsec:SCB}
\paragraph{}
As described in \cite{Nirschl:2004pa,Dolan:2004mu} (see also \cite{Bissi:2015qoa,Doobary:2015gia}), the solutions to the quadratic superconformal Casimir equation are classified by the representations of the exchanged operators under the superconformal symmetry group $PSU(2, 2|4)$.
In this paper, we focus on the case that the exchanged state belongs to {the so-called long multiplet with $\Delta\,, J$ and the $SU(4)$ representation $[s,l-s,s]$\,\cite{Dobrev:1985qv}}.

\medskip

The solution to the superconformal Casimir equation associated with a long multiplet has the following form \cite{Nirschl:2004pa,Dolan:2004mu,Bissi:2015qoa,Doobary:2015gia}:
\begin{align}
\cG_{\Delta,J,l,s}^{a,b}(z_i,\alpha_i)\Bigl|_{\text{Long}}&=\cF(z_i,\alpha_i)H^{a,b}_{\Delta,J,l,s}(z_i,\alpha_i)\,,
\end{align}
where the explicit overall factor $\cF(z_i,\alpha_i)$ is given by
\begin{align}
\cF(z_i,\alpha_i)&=\frac{(z_1-\alpha_1)(z_1-\alpha_2)(z_2-\alpha_1)(z_2-\alpha_2)}{(z_1z_2)^2(\alpha_1\alpha_2)^2}
\,.
\end{align}
{If we perform the similarity transformation on $\cD_{\mathfrak{b}}^{(a,b)}$ and $\cD_{\mathfrak{f}}$ using $\cF(z_i,\alpha_i)$, we can simplify the Casimir operator in \eqref{eq:superCsimireq} into:}
\begin{align}
&\cF(z_i,\alpha_i)^{-1}\left(\cD_{\mathfrak{b}}^{(a,b)}+\cD_{\mathfrak{f}}\right)\cF(z_i,\alpha_i)=
\cD_{\mathfrak{b}}^{(a,b)}\,.
\end{align}
Therefore, the reduced superconformal block $H^{a,b}_{\Delta,J,l,s}(z_i,\alpha_i)$ satisfies the quadratic Casimir equation for $SO(2,4)\times SO(6)$\,,
\begin{align}
\left(\check{\cD}_{4}^{(a,b)}-\hat{\cD}_{4}^{(a,b)}\right)H^{a,b}_{\Delta,J,l,s}(z_i,\alpha_i)
=\frac{1}{2}C_{\Delta,J,l,s}H^{a,b}_{\Delta,J,l,s}(z_i,\alpha_i)\,.
\label{eq:long-Casimireq}
\end{align}
This fact implies that $H^{a,b}_{\Delta,J,l,s}(z_i,\alpha_i)$ can be further decomposed into
\begin{align}\label{H-block}
H^{a,b}_{\Delta,J,l,s}(z_i,\alpha_i)&\equiv
\check{G}_{\Delta+4,J}^{a,b}(z_i)\hat{G}^{a,b}_{l,s}(\alpha_i)\,.
\end{align}
The conformal symmetry dependent part $\check{G}_{\Delta+4,J}^{a,b}(z_i)$ is the usual four dimensional conformal block given by
\begin{align}
\check{G}_{\Delta,J}^{a,b}(z_i)&=\frac{z_1z_2}{z_1-z_2}\left[k^{a,b}_{\frac{\Delta-J}{2}-1}(z_1)k^{a,b}_{\frac{\Delta+J}{2}}(z_2)-k^{a,b}_{\frac{\Delta-J}{2}-1}(z_2)k^{a,b}_{\frac{\Delta+J}{2}}(z_1)\right]\,,\\
k_{\rho}^{a,b}(z)&\equiv
z^\rho{}_2F_1\left(\rho+a,\rho+b~;2\rho~;z\right)\,.
\label{def:k-hyper}
\end{align}
On the other hand, the $R$-symmetry dependent part: $\hat{G}^{a,b}_{l,s}(\alpha_i)$ is called $R$-symmetry block, and the explicit expression is \cite{Nirschl:2004pa,Bissi:2015qoa,Doobary:2015gia}
\begin{align}
\hat{G}_{l,s}^{a,b}(\sigma,\tau)\propto\, (\alpha_1\alpha_2)^{a}\,P_{\frac{l+s}{2}+a,\frac{l-s}{2}+a}^{(-a+b,-a-b)}(\hat{w}_1,\hat{w}_2)\,.
\label{eeq:R-sym-block}
\end{align}
Here, $P^{(\alpha,\beta)}_{m,n}(\hat{w}_1,\hat{w}_2)$ can be expressed in terms of the Jacobi polynomial
\begin{align}
P^{(\alpha,\beta)}_{m,n}(\hat{w}_1,\hat{w}_2)&=
\frac{P_{m+1}^{(\alpha,\beta)}(\hat{w}_1)P_{n}^{(\alpha,\beta)}(\hat{w}_2)-P_{n}^{(\alpha,\beta)}(\hat{w}_1)P_{m+1}^{(\alpha,\beta)}(\hat{w}_2)}{\hat{w}_1-\hat{w}_2}\,,\\
P_{n}^{(\alpha,\beta)}(\hat{w})&=\frac{(\alpha+1)_{n}}{n!} {}_2F_1\left(-n,1+\alpha+\beta+n;\alpha+1;\frac{1-\hat{w}}{2}\right)\,,
\label{def:Jacobi-4d}
\end{align}
where $\hat{w}_{1}$ and $\hat{w}_{2}$ are defined through:
\begin{align}
    \alpha_i=\frac{2}{1-\hat{w}_i}\,.\label{def:a-w}
\end{align}
The solution (\ref{eeq:R-sym-block}) to the quadratic Casimir equation is obtained by specifying the boundary condition,
\begin{align}
\begin{split}
&\hat{G}_{l,s}^{\hat{a},\hat{b}}(\sigma,\tau)\sim (\alpha_1 \alpha_2)^{-\frac{1}{2}l}\,C_{s}^{(1)}\left(\frac{\alpha_1+\alpha_2}{2\sqrt{\alpha_1\alpha_2}}\right)+\cO\left((\alpha_1\alpha_2)^{-\frac{l}{2}+1}\right) \,,\\
&\alpha_1 \alpha_2\to 0\,,\qquad \frac{\alpha_1+\alpha_2}{2\sqrt{\alpha_1\alpha_2}}=\text{fixed}\,,
\label{eq:leading-GR}
\end{split}
\end{align}
where $C_{n}^{(m)}(x)$ is the Gegenbauer polynomial.
The holographic dual to the global conformal block has been constructed by considering the so-called geodesic Witten diagram in five dimensional AdS space \cite{Hijano:2015zsa},
our main goal in this paper is to extend this result and construct a holographic dual configuration along the spherical space which reproduce to the $R$-symmetry block (\ref{eeq:R-sym-block}), hence the full superconformal block $H^{a,b}_{\Delta,J,l,s}(z_i,\alpha_i)$ \eqref{H-block}.

\section{$R$-symmetry block from Harmonic Analysis on $S^{d+1}$}\label{Sec:Harmonic}
\paragraph{}
In this section, we would like to generalize the widely used embedding space formalism for AdS space to the spherical geometry to construct the holographic dual to the $R$-symmetry block, for an introduction, see \cite{Rychkov:2016iqz}.
In particular, we will present an generalization of the AdS split representations to the spherical harmonic function and the formulas of three-point functions on $S^{d+1}$\,.
As we will see, these are ingredients for constructing a holographic dual to the $R$-symmetry block. Moreover, we should stress here that while our main focus will be $AdS_5\times S^5$ geometry, our analysis however is valid for the sphere of arbitrary dimensions $S^{d+1}$.

\subsection{Embedding formalism for $S^{d+1}$}
\paragraph{}
Similar to the AdS case, it is convenient to employ the embedding formalism to describe functions on $S^{d+1}$ for the construction of a holographic dual to the $R$-symmetry block.
{The embedding formalism for sphere can mostly be constructed by performing analytical continuations of the physical quantities in AdS one.}
Therefore, here we only summarize the major differences between the AdS and the spherical embedding formalism.
For more details of the spherical case, please see appendix \ref{App:embedding}.

\subsubsection*{Embedding space coordinates}
\paragraph{}
In the embedding formalism,  $AdS_{d+1}$ and $S^{d+1}$ are realized as the hyper-surfaces in $\mathbb{R}^{2,d}$ and $\mathbb{R}^{d+2}$\,, respectively:
\begin{align}
\text{AdS}&:\quad\chX^2\equiv\eta_{\chA\chB}\chX^{\chA}\chX^{\chB}=-1\,,\qquad \chA=1,\dots, d+2\,,\\
\text{sphere}&:\quad\hX^2\equiv\delta_{\hA\hB}\hX^{\hA}\hX^{\hB}=+1\,,\qquad \hA=1,\dots, d+2\,,
\end{align}
where $\eta_{\chA\chB}=(-1,-1,1,\dots,1)$ and $\delta_{\hA\hB}$ is the Kronecker delta.
The boundary of $AdS_{d+1}$ is expressed as a null cone surface in $\mathbb{R}^{2,d}$\,,
\begin{align}
\text{AdS}:\quad P^{\chA}\in \mathbb{R}^{2,d}\,,\qquad P\cdot P\equiv \eta_{\chA\chB}P^{\chA}P^{\chB}=0\,,
\label{def:c-cone}
\end{align}
which is how we realize $d$-dimensional Minkowski space $\mathbb{R}^{1, d-1}$ after taking the Poincare slice.
\paragraph{}
On the other hand, the sphere does not have the boundary, but let us recall that in the previous section, the $R$-symmetry cross ratios are constructed from the complex null vectors $T^{\hat{A}}$ associated with the half-BPS operators. 
Therefore, we will regard the complex null cone as the spherical counterpart of the AdS boundary \cite{Bargmann1977}: 
\begin{align}
\text{sphere}:\quad T^{\hA}\in \mathbb{C}^{d+2}\,,\qquad T\cdot T\equiv \delta_{\hA\hB}T^{\hA}T^{\hB}=0\,.
\label{def:cnullcone}
\end{align}
Note the complex null cone is invariant under the (complex) dilatation transformation,
\begin{align}
    T^{\hat{A}} \sim \la\,T^{\hat{A}}\,,\qquad \la \in \mathbb{C}^{\times}\equiv \mathbb{C}\backslash \{0\}\,,\label{comp-dil}
\end{align}
as $T^{\hat{A}}$ is now complex.
{We should note that the main difference between the AdS-space and the spherical embedding formalisms is that there does not exist a notion of a geodesic interpolating between the bulk space $S^{d+1}$ to the complex null cone in contrast with the AdS case.}
This distinction plays an important role to construct a holographic dual of the $R$-symmetry block.

\subsubsection*{Bulk-to-boundary propagator}
\paragraph{}
Having clarified what we meant by the ``boundary'' and ``bulk' coordinates in the spherical embedding space formalism,
let us next consider the counterparts of the AdS bulk-to-boundary propagators.
The AdS scalar bulk-to-boundary propagator is given by
\begin{align}
\text{AdS}:\quad K_{\Delta,0}^{\sfa}(\chX;P)&=\chC_{\Delta,0}(\chX\cdot P)^{-\Delta}\,,
\end{align}
where $\chC_{\Delta,0}$ is an overall normalization constant.
This function satisfies the quadratic Casimir equation for $SO(2,d)$ with the eigenvalue $\Delta(\Delta-d)$\,.
The eigenvalue can be mapped to the spherical one by a direct replacement $\Delta\leftrightarrow -l$.
Therefore, through replacements of embedding space coordinates $\chX_{\chA}\leftrightarrow \hX_{\hA}$\,, $P_{\chA}\leftrightarrow T_{\hA}$ and $\Delta\leftrightarrow -l$, a natural counterpart of the AdS scalar bulk-to-boundary propagator is obtained as
\begin{align}
\text{sphere}:\quad K_{l,0}^{\sfs}(\hX;T)&=\hC_{l,0}(\hX\cdot T)^{l}\,,
\label{def:shf}
\end{align}
where $\hC_{l,0}$ is an overall normalization constant remained to be fixed.
The null condition (\ref{def:cnullcone}) is required in order to describe the symmetric traceless tensors which belong to the irreducible representations of $SO(d+2)$\,. 
This polynomial is known as the scalar spherical harmonic (SSH) function.
The space of such homogeneous polynomials has been extensively discussed in \cite{Bargmann1977}.
When $d=4$\,, the SSH function corresponds to the $SU(4)$ symmetric traceless representation labelled by the Dynkin label $[0,l,0]$.

\paragraph{}

The generalization to the spinning case can easily be done.
In the AdS case, if we introduce the additional null vectors $W_{\chA}\,, Z_{\chA}\in\mathbb{R}^{2,d}$ satisfing
\begin{align}
  \text{AdS}:\quad    \check{X}\cdot W=0\,,\qquad P\cdot Z=0\,,
\end{align}
the AdS spinning bulk-to-boundary propagator is given by
\begin{align}
\begin{split}
  \text{AdS}:\quad  K_{\Delta,J}^{\sfa}(\chX,W;P,Z)&=\chC_{\Delta,J}(\chX\cdot \check{\sfC}\cdot W)^{J}(\chX\cdot P)^{-\Delta-J}\,,\\
\check{\sfC}_{\chA\chB}&=P_{\chA}Z_{\chB}-Z_{\chA}P_{\chB}\,,
\end{split}
\end{align}
where $\chC_{\Delta,J}$ is an overall normalization constant.
As in the AdS case, we can further introduce the additional auxiliary complex null vectors $U^{\hA}\,, R^{\hA}\in \mathbb{C}^{d+2}$ satisfying 
\begin{align}
 \text{sphere}:\quad   \hat{X}\cdot U=0\,,\qquad T\cdot R=0\,.
\end{align}
The spherical spinning bulk-to-boundary propagators are given by 
\begin{align}
\begin{split}
 \text{sphere}:\quad K_{l,s}^{\sfs}(\hX,U;T,R)&=\hC_{l,s}(\hX\cdot \hsfC\cdot U)^{s}(\hX\cdot T)^{l-s}\,,\\
 \hsfC_{\hA\hB}&=T_{\hA}R_{\hB}-R_{\hA}T_{\hB}\,,
 \label{def:tsh}
 \end{split}
\end{align}
where the parameter $s$ now plays the role of ``spin'' and $\hC_{l,s}$ is an overall normalization constant.
When $d=4$\,, the tensor harmonic function corresponds to the $SU(4)$ representation labelled by the Dynkin label $[s,l-s,s]$.

\subsection{Split representations of harmonic functions}\label{subsec:split}
\paragraph{}
Now we would like to construct the spherical harmonic functions using the spherical bulk to boundary propagators we just listed, this can be regarded as a generalization of split representation of AdS harmonic function \cite{Costa:2014kfa}.
First of all, let us consider a harmonic function on the sphere $S^{d+1}$ which depends on two bulk points $\hX_1\,,\hX_2$ and two polarization {null vectors} $U_1\,, U_2$\,.
The function satisfies
\begin{align}
(\hat{\nabla}^2_1+l(l+d)-s)\hat{\Omega}_{l,s}(\hX_1,U_1;\hX_2,U_2)=0\,,
\label{def:tharmonic-omega-eq}
\end{align}
where $\hat{\nabla}_{1,\hA}$ is the covariant derivative with respect to $\hat{X}_1\,, U_1$ defined in (\ref{def:covariant-der}).
For general $s$, a solution to the equation (\ref{def:tharmonic-omega-eq}) can be constructed by generalizing the AdS split representation \cite{Costa:2014kfa} to  $\hat{\Omega}_{l,s}(\hX_1,U_1;\hX_2,U_2)$, as we will do in this subsection.

\subsubsection*{Scalar case}
\paragraph{}
As a warm up, let us consider the scalar $s=0$ case.
In this case, the harmonic function $\hat{\Omega}_{l,0}(\hX_1;\hX_2)$ only depends on $z\equiv\hat{X}_{12}\equiv \hat{X}_1\cdot \hat{X}_2$ from the $SO(d+2)$ symmetry.
The equation of motion (\ref{def:tharmonic-omega-eq}) becomes 
\begin{align}
    \left[(1-z^2)\frac{d^2}{dz^2}-(d+1)z\frac{d}{dz}+l(l+d)\right]\hat{\Omega}_{l,0}(z)=0\,,
\end{align}
and a solution to this equation is given by the Gegenbauer polynomial
\begin{align}
\hat{\Omega}_{l,0}(\hX_1;\hX_2)\propto\,C_{l}^{\left(h\right)}(z)\,,\qquad h\equiv \frac{d}{2}\,.\label{eq:gegen-sol}
\end{align}
This solution can be reproduced by a split representation of $\hat{\Omega}_{l,0}(\hX_1;\hX_2)$,
\begin{align}
\hat{\Omega}_{l,0}(\hX_1;\hX_2)=\int_{\mathbb{C}^{d+2}} D^{d}T\,K_{l,0}^{\sfs}(\hX_1;T)K_{\tilde{l},0}^{\sfs}(\hX_2;T)\,.
\label{def:sharmonic-split}
\end{align}
Here the integration measure is defined by
\begin{align}
\int_{\mathbb{C}^{d+2}} D^{d}T&\equiv\int_{\mathbb{C}^{d+2}}  \frac{\dd^{d+2}T\,\delta(T\cdot T)}{{\rm Vol}(\mathbb{C}^{\times})}\no\\
&\equiv\int_{\mathbb{R}^{d+2}\times \mathbb{R}^{d+2}} \frac{\dd^{d+2} T_{\sfr}\,\dd^{d+2} T_{\sfi}}{{\rm Vol}(\mathbb{C}^{\times})}
\delta(T_{\sfr}^2-T_{\sfi}^2)\delta(2T_{\sfr}\cdot T_{\sfi})\,,
\label{eq:harmonic-split1}
\end{align}
where {the symbol ${\rm Vol}(\mathbb{C}^{\times})$ expresses the volume for the complexified dilatation transformation (\ref{comp-dil}), and} $T_{\mathsf{r}}$ and $T_{\mathsf{i}}$ are the real and imaginary parts of $T$\,, respectively.
Here, $\tilde{l}$ is a negative integer defined as: 
\begin{align}
\tilde{l}\equiv -d-l\,,\label{eq:tilde-l}
\end{align}
which is a counterpart of the shadow scaling dimension $\tilde{\Delta}=d-\Delta$.
{If we define a conformal integral in analogy to the AdS case, the definition (\ref{eq:tilde-l}) of $\tilde{l}$ can be fixed by requiring the invariance of the integral (\ref{eq:harmonic-split1}) under the complexified dilatation transformation (\ref{comp-dil})
}\footnote{Another motivation to use $\tilde{l}$ is that the the Gegenbauer function satisfies the following relation \cite{Gegenbauerid}:
\begin{align}
    C^{(h)}_{l}(z)=\frac{\sin\pi l}{\sin\pi\tilde{l}}C^{(h)}_{\tilde{l}}(z)\,.
\end{align}}.
In fact, the eigenvalue of the quadratic Casimir operator is invariant under the replacement $l\to \tilde{l}$\,.
Furthermore, the use of the bulk-to-boundary propagator with $\tilde{l}$ in the split representation guarantees that the integral (\ref{def:sharmonic-split}) is invariant under the action of $SO(d+2)$\,.
The integral (\ref{def:sharmonic-split}) is performed in appendix \ref{subsec:HfunctionS}, and the final result is again proportional to the Gegenbauer polynomial
\begin{align}
\hat{\Omega}_{l,0}(\hX_1;\hX_2)=\frac{{\rm Vol}(S^{d})\hC_{l,0}\hC_{\tilde{l},0}}{C_{l}^{\left(h\right)}(1)}
C_{l}^{\left(h\right)}(z)\,,
\label{eq;scalar-split-result}
\end{align}
where the symbol ${\rm Vol}(S^{d})$ expresses the volume of the $d$-dimensional sphere $S^{d}$\,.
In this way, the split representation (\ref{eq:harmonic-split1}) reproduces (\ref{eq:gegen-sol}).

\subsubsection*{Tensor case}
\paragraph{}
The extension to the tensor case is straightforward. By using the tensor bulk-to-boundary propagator (\ref{def:tsh}), we can construct the split representation of $\hat{\Omega}_{l,s}$\,:
\begin{align}
\hat{\Omega}_{l,s}(\hX_1,U_1;\hX_2,U_2)=\frac{1}{\left(h-1\right)_{s}s!}\int_{\mathbb{C}^{d+2}} D^{d}T\,K_{l,s}^{\sfs}(\hX_1,U_1;T,\hat{D}_{R})K_{\tilde{l},s}^{\sfs}(\hX_2,U_2;T,R)\,,
\label{def:tharmonic-split}
\end{align}
where $(x)_n$ is the 
pochhammer symbol and the differential operator $\hat{D}_{R}$ responsible for index contraction is defined in (\ref{def:DR}).
As shown in appendix \ref{subsec:HfunctionS}, the harmonic function (\ref{def:tharmonic-split}) after the integration takes the form:
\begin{align}
    \hat{\Omega}_{l,s}(\hX_1,U_1;\hX_2,U_2)&=\sum_{r=0}^{s}U_{12}^{s-r}(U_1\cdot \hat{X}_2)^r(U_2\cdot \hat{X}_1)^r\,g_{l,s}^{(r)}(z)\,,
    \label{eq:Omega-form}
\end{align}
and by construction, satisfies the equation of motion (\ref{def:tharmonic-omega-eq}), as the bulk to boundary propagators in the integrand do.

\paragraph{}

Let us see the $s=1$ case in more details.
The equation of motion (\ref{def:tharmonic-omega-eq}) can be rewritten as
\begin{align}
    0&=U_{12}\biggl[\hat{\nabla}^2_1g_{l,1}^{(0)}(z)+\left(l(l+d)-2\right)g_{l,1}^{(0)}(z)-2z\,g_{l,1}^{(1)}(z)\biggr]+(U_1\cdot \hat{X}_2)(U_2\cdot \hat{X}_1)\times\no\\
    &\quad \times\biggl[\hat{\nabla}^2_1g_{l,1}^{(1)}(z)-4z\,\frac{d}{dz}g_{l,1}^{(1)}(z)+(l(l+d)-(d+3))g_{l,1}^{(1)}(z)-2 g_{l,1}^{(0)}(z)\biggr]\,,
    \label{eq:EOM-spin1}
\end{align}
where the action of $\hat{\nabla}^2_1$ on any function $f(z)$ is given by:
\begin{align}
\hat{\nabla}^2_1f(z)=\left\{(1-z^2)\frac{d^2}{dz^2}-(d+1)z\frac{d}{dz}\right\} f(z)\,.
\end{align}
{{The integral (\ref{def:tharmonic-split})} can be performed, as done in appendix \ref{subsec:HfunctionS} following the procedure similar to the one described in appendix C of \cite{Costa:2014kfa}.}
By performing the integral (\ref{def:tharmonic-split}), we obtain
\begin{align}
    \hat{\Omega}_{l,1}(\hX_1,U_1;\hX_2,U_2)&=U_{12}\,g_{l,1}^{(0)}(z)+(U_1\cdot \hat{X}_2)(U_2\cdot \hat{X}_1)\,g_{l,1}^{(1)}(z)\,,   \label{eq:harmonic-spin1}
    \end{align}
where $g_{l,1}^{(0)}(z)$ and $g_{l,1}^{(1)}(z)$ are given by
\begin{align}
\begin{split}
   g_{l,1}^{(0)}(z)&=\mathcal{N}\biggl(\frac{C_{l}^{(h)}(z)}{C_{l}^{(h)}(1)}-\frac{z}{2h+1}\,\frac{C_{l-1}^{(h+1)}(z)}{C_{l-1}^{(h+1)}(1)}\biggr)\,,\\
   g_{l,1}^{(1)}(z)&=-\mathcal{N}\biggl(\frac{2(h+1)(l-1)(2h+l+1)z}{(2h+1)_{3}}\,\frac{C_{l-2}^{(h+2)}(z)}{C_{l-2}^{(h+2)}(1)}+\frac{(2h)^2}{l(2h+l)}\frac{C_{l-1}^{(h+1)}(z)}{C_{l}^{(h)}(1)}\biggr)\,,\\
   \mathcal{N}&=\text{Vol}(S^{d})\hC_{l,s}\hC_{\tilde{l},s}\,.\label{eq:harmonic-spin1-g}
   \end{split}
\end{align}
We can check that (\ref{eq:harmonic-spin1}) and  (\ref{eq:harmonic-spin1-g}) together satisfy the equation of motion for $s=1$ (\ref{eq:EOM-spin1}).

\paragraph{}

In general, performing the integral in (\ref{def:tharmonic-split}) requires rather cumbersome calculations.
However, when we take a special choice of the polarization vectors $U_1\,, U_2$ such that $\hat{X}_2\cdot U_1=0=\hat{X}_1\cdot U_2$\,, we can easily perform the integral and obtain
\begin{align}
&\hat{\Omega}_{l,s}(\hX_1,U_1;\hX_2,U_2)
=U_{12}^{s}\,g^{(0)}_{l,s}(z)\,,\\
&g^{(0)}_{l,s}(z)=\text{Vol}(S^{d})\hC_{l,s}\hC_{\tilde{l},s}
\sum_{r=0}^{s}\frac{(-1)^rs!r!\Gamma(2h)(h)_{r}}
{(s-r)!\Gamma(2(h+r))}\frac{C_{l-r}^{(h+r)}(z)}{C_{l-r}^{(h+r)}(1)}\no\\
&\qquad \qquad\times 
\,(h-1+s)_{-r}\,C_{r}^{(h-1+s-r)}(z)\,.
\label{eq:split-tensor-result}
\end{align}
{The details of the derivation for (\ref{eq:split-tensor-result}) are given in appendix \ref{subsec:HfunctionS}\,.
As a consistency check, we can see that the expression (\ref{eq:split-tensor-result}) reproduces the $s=0$ case (\ref{eq;scalar-split-result}) and $g_{l,1}^{(0)}(z)$ in (\ref{eq:harmonic-spin1-g}).
Note that the expression (\ref{eq:split-tensor-result}) does not solve the equations of motion (\ref{eq:EOM-spin1}), for $\hat{\Omega}_{l,s}(\hX_1,U_1;\hX_2,U_2)$ to satisfy the equation of motion (\ref{def:tharmonic-omega-eq}), not only a leading term $g^{(0)}_{l,s}(z)$ but also a higher-order terms $g^{(r)}_{l,s}(z)\,(1\leq r\leq s)$ are needed (For the $s=1$ case, see (\ref{eq:EOM-spin1})). }

\subsection{Three-point functions}\label{subsec-3pt}
\paragraph{}
Next, we will consider the simplest three-point functions involving  only three spherical bulk-to-boundary propagators, this is the spherical analogue of three-point contact Witten diagram in AdS space, and serves as the building block for higher point correlation functions.

\subsubsection{Scalar-scalar-scalar case}
\paragraph{}
Let us first see three-point functions involving three scalar bulk-to-boundary propagators, which are given by
\begin{align}
\hA_{l_1l_2l_3}(T_i)\equiv \int_{S^{d+1}}\dd^{d+2}{\bf \hX}\,
 K_{l_1,0}^{\sfs}(\hX;T_1)K_{l_2,0}^{\sfs}(\hX;T_2)K_{l_3,0}^{\sfs}(\hX;T_3)\,,
\label{eq:YYY-int}
\end{align}
where the integration measure is defined in (\ref{eq:measure-S}).
Temporarily, we assume that all quantum numbers $l_i\,(i=1,2,3)$ are positive integers, {and we will analytically continue them in the final expression (\ref{eq:3pt-overallconst-gamma})} obtained after the integration.
As shown in appendix \ref{app:SHfun}, the above integral can be evaluated as
\begin{align}
\hA_{l_1l_2l_3}(T_i)&=\hsfA_{l_1l_2l_3}
T_{23}^{\alpha_{231}}T_{31}^{\alpha_{312}}T_{12}^{\alpha_{123}}\,,
\label{eq:YYY-formula}
\end{align}
where the overall constant $\hsfA_{l_1l_2l_3}$ and $\alpha_{ijk}$ are given by
\begin{align}
\hsfA_{l_1l_2l_3}&=\text{Vol}(S^{d+1})\frac{(-1)^{L_{123}}\Ga\left(h+1\right)}{\Ga\left(h+1+L_{123}\right)}
\frac{l_1!l_2!l_3!}{\alpha_{231}!\alpha_{312}!\alpha_{123}!}
\hC_{l_1,0}\hC_{l_2,0}\hC_{l_3,0}\,,\label{eq:3pt-overallconst}\\
L_{ijk}&=\frac{l_i+l_{j}+l_{k}}{2}\,,\qquad
\alpha_{ijk}=\frac{l_i+l_{j}-l_{k}}{2}\,.
\end{align}
The above formula (\ref{eq:YYY-formula}) can also be expressed in terms of the anti-symmetric $C$-tensor \cite{Lee:1998bxa,Arutyunov:1999en}.
From the overall coefficient $\hsfA_{l_1l_2l_3}$, the integration (\ref{eq:YYY-int}) with positive integers $l_i$ does not vanish only if each $\alpha_{ijk}$ satisfies
\begin{align}
\alpha_{123}\,, \quad \alpha_{321}\,,\quad \alpha_{231}\in \mathbb{Z}_{\geq0}\,.
\label{eq:YYY-int-con}
\end{align}

\medskip

In the next subsection, we will evaluate an integral representation of the $R$-symmetry block by employing the split representation (\ref{def:tharmonic-split}) of the spherical harmonic function.
For this purpose, it is useful to rewrite the overall constant (\ref{eq:3pt-overallconst}) in terms of gamma functions as
\begin{align}
\hsfA_{l_1l_2l_3}&=\text{Vol}(S^{d+1})\frac{(-1)^{L_{123}}\Ga\left(h+1\right)}{\Ga\left(h+1+L_{123}\right)}
\frac{\Gamma(l_1+1)\Gamma(l_2+1)\Gamma(l_3+1)}{\Gamma(\alpha_{231}+1)\Gamma(\alpha_{312}+1)\Gamma(\alpha_{123}+1)}
\hC_{l_1,0}\hC_{l_2,0}\hC_{l_3,0}\,.
\label{eq:3pt-overallconst-gamma}
\end{align}
This expression can be analytically continued to complex values $l_i$ except for the poles $l_i=-n-1\,, n\in\mathbb{Z}_{\geq0}$ of the gamma functions appearing in the numerator.
Note that the Gamma functions appearing in (\ref{eq:3pt-overallconst-gamma}) are consistent with (127) in \cite{Costa:2014kfa} with the replacement $-l_i = \Delta_i$ by using the reflection formula $\Gamma(z)\Gamma(1-z)=\frac{\pi}{\sin \pi z}$, we obtain:
\begin{align}
\hsfA_{l_1l_2l_3}&=2(-1)^{L_{123}}\sin\pi\left(h+L_{123}\right)\frac{\sin\pi\alpha_{231}\sin\pi\alpha_{312}\sin\pi\alpha_{123}}{\sin\pi l_1\sin\pi l_2\sin\pi l_3}\no\\
&\quad \times \pi^{\frac{d}{2}}\Ga\left(-h-L_{123}\right)
\frac{\Gamma(-\alpha_{231})\Gamma(-\alpha_{312})\Gamma(-\alpha_{123})
}{\Gamma(-l_1)\Gamma(-l_2)\Gamma(-l_3)}
\hC_{l_1,0}\hC_{l_2,0}\hC_{l_3,0}\,.
\end{align}

\subsubsection{Scalar-scalar-tensor case}
\paragraph{}
Next, let us consider the three-point functions involving tensor spherical bulk-to-boundary propagator.
Here we will consider the scalar-scalar-tensor case given by
\begin{align}
\hA_{l_1l_2l_3;0,0,s_3}
&\equiv\int_{S^{d+1}}\dd^{d+2}{\bf \hX}
 \left(\cY_3^{s_3}K_{l_1,0}^{\sfs}(\hX;T_1)\right)
K_{l_2,0}^{\sfs}(\hX;T_2)K_{l_3,s_3}^{\sfs}(\hX,U_3;T_3,R_3)\,,
\label{eq:SST-int}
\end{align}
where $\cY_3\equiv \partial_{U_3}\cdot \partial_{\hX}$ and we assume $l_2\geq l_1 \geq s_3$\,.
{We can also consider other three-point functions that the derivatives with respect to $\hX^{\hA}$ act on either of $K_{l_1,0}^{\sfs}(\hX;T_1)$ and $K_{l_2,0}^{\sfs}(\hX;T_2)$\,. By performing partial integrals, these vertices are reduced to the form (\ref{eq:SST-int}) because $K_{l_3,s_3}^{\sfs}(\hX,U_3;T_3,R_3)$ is divergence free, and the sphere does not have the boundary hence the additional boundary contributions.} 
The $s=1,2$ cases have been computed in \cite{Arutyunov:1999en}.

\medskip

This integral can be easily computed by using the relation (\ref{eq:spin-scalar-s}), and the final expression for (\ref{eq:SST-int}) can be written as:
\begin{align}
\hA_{l_1l_2l_3;0,0,s_3}(T_i,R_3)
&=\frac{s_3!(l_1-s_3+1)_{s_3}\hat{C}_{l_3,s_3}\hat{C}_{l_1,0}}{(l_3-s_3+1)_{s_3}\hat{C}_{l_3,0}\hat{C}_{l_1-s_3,0} } \hcD_{T_3}(T_1,R_3)^{s_3}\hA_{l_1-s_3,l_2,l_3}(T_i)\no\\
&=  \hsfA_{l_1l_2l_3;0,0,s_3}\hat{\mathsf{V}}_{3,12}^{s_3}T_{23}^{\alpha_{231}-\frac{s_3}{2}}T_{31}^{\alpha_{312}-\frac{s_3}{2}}T_{12}^{\alpha_{123}+\frac{s_3}{2}}\,,
\label{eq:SST-3pt}
\end{align}
where the differential operator $\hcD_{T_3}(T_1,R_3)$ is defined in (\ref{cDT-def}), and we have introduced 
\begin{align}
\hat{\mathsf{V}}_{i,jk}=-\frac{T_j\cdot \hsfC_{i}\cdot T_{k}}{T_{j}\cdot T_{k}}\,.
\end{align}
In the second equation, we have used the fact
\begin{align}
\hcD_{T_3}(T_1,R_3)\hat{\mathsf{V}}_{3,12}=0\,.
\end{align}
The overall constant $\hsfA_{l_1l_2l_3;0,0,s_3}$ is given by
\begin{align}
 \hsfA_{l_1l_2l_3;0,0,s_3}&=\frac{s_3!(l_1-s_3+1)_{s_3}\Gamma\left(\alpha_{231}+\frac{s_3}{2}+1\right)\hat{C}_{l_1,0}\hat{C}_{l_3,s_3}     }{ (l_3-s_3+1)_{s_3}\Gamma\left(\alpha_{231}-\frac{s_3}{2}+1\right) \hat{C}_{l_1-s_3,0}\hat{C}_{l_3,0}   } \hsfA_{l_1-s_3l_2l_3}\no\\
 &=\text{Vol}(S^{d+1})\frac{(-1)^{L_{123}-\frac{s_3}{2}}\Ga\left(h+1\right)}{\Ga\left(h+1+L_{123}-\frac{s_3}{2}\right)}
\frac{s_3!\Gamma(l_1+1)\Gamma(l_2+1)\Gamma(l_3-s_3+1)\hat{C}_{l_1,0}\hat{C}_{l_2,0}\hat{C}_{l_3,s_3}}{\Gamma(\alpha_{231}-\frac{s_3}{2}+1)\Gamma(\alpha_{312}-\frac{s_3}{2}+1)\Gamma(\alpha_{123}-\frac{s_3}{2}+1)}
\,.\label{eq:three-const-spin}
\end{align}
{The Gamma function factors of the three-point function above are also consistent with (131) in \cite{Costa:2014kfa}.}

\subsection{Holographic dual to the $R$-symmetry block }\label{sec:R-symb}
\paragraph{}
Finally, let us use the three-point functions we just obtained to construct a holographic dual to the $R$-symmetry block.

\medskip

For this purpose, we first give a holographic dual to the $R$-symmetry partial waves.
Suppose that the external and exchanged states transform in the $R$-symmetry representations $[s_{i},l_{i}-s_{i},s_{i}]$ and $[s,l-s,s]$, respectively.
Then a possible holographic dual of the $R$-symmetry partial waves is the following four-point function of $K_{l_i,s_i}^{\sfs}(\hX_i,U_i;T_i,R_i)$:
\begin{align}
\hW_{l,s}^{\{l_i,s_i\}}(T_i,R_i)&\equiv \frac{1}{\pi^{h}\,\hsfA_{l_1l_2l;s_1,s_2,s}\,\hsfA_{\tilde{l}l_3l_4;s,s_3,s_4} }\int_{S^{d+1}} \dd^{d+2}{\bf \hX}\,\int_{S^{d+1}} \dd^{d+2}{\bf \hX'}\no\\
&\times \cY^{s}_{5}\cY^{s}_{6}K_{l_1,s_1}^{\sfs}(\hX_1,U_1;T_1,R_1)K_{l_2,s_2}^{\sfs}(\hX_2,U_2;T_2,R_2)\hat{\Omega}_{l,s}(\hX_5,U_5;\hX_6,U_6)\no\\
&\qquad\times K_{l_3,s_3}^{\sfs}(\hX_3,U_3;T_3,R_3)K_{l_4,s_4}^{\sfs}(\hX_4,U_4;T_4,R_4)\Bigl\lvert_{\hX_{1,2,5}=\hX\,,\hX_{3,4,6}=\hX_{1}}\,,
\label{eq:four-R-int}
\end{align}
where $\cY_{5}\equiv \partial_{U_5}\cdot \partial_{\hX^1}$ and $\cY_{6}\equiv \partial_{U_6}\cdot \partial_{\hX^3}$\,.
For $l_i=s_i=0$, this four-point function is an eigenfunction of the quadratic Casimir equation for $SO(d+2)$ by construction.
{In particular, we do not restrict the interaction vertices to move along the geodesics on the sphere as the integration regions to describe the $R$-symmetry block holographically.
This follows from the fact that there does not exist a notion of geodesics interpolating the bulk space $S^{d+1}$ to the complex null cone (\ref{def:cnullcone}), hence integrating over the entire $S^{d+1}$ generates the only $SO(d+2)$-preserving configuration.}

\paragraph{}

{As we will see, after performing integrals in (\ref{eq:four-R-int}), the $R$-symmetry partial wave (\ref{eq:four-R-int}) has a form which is a linear combination of the $R$-symmetry block (\ref{eq:R-block-series}) and its shadow (\ref{eq:R-block-series-sh}), as in the AdS case.}
Remarkably, we will observe that the shadow part vanishes after imposing the selection rule (\ref{eq:selection-R-sym}) on the exchanged operator after taking an appropriate normalization (\ref{eq:W-normalize}).
In this way, our $R$-symmetry partial wave (\ref{eq:four-R-int}) also describes the $R$-symmetry block. This situation is different from the AdS case.

\paragraph{}

Let us give a brief sketch on how we perform the integrals of (\ref{eq:four-R-int}) with $s_{i}=0$ which is the same method applied in \cite{Dolan:2011dv}.
For this purpose, we first use the split representation (\ref{def:tharmonic-split}) of the harmonic function $\hat{\Omega}_{l,s}$ and the formula (\ref{eq:SST-3pt}) of three-point functions.
Then the $R$-symmetry partial wave becomes: 
\begin{align}
\hW_{l,s}^{\{l_i,0\}}&=\frac{1}{\pi^{h}\,\hsfA_{l_1l_2l;0,0,s}\,\hsfA_{\tilde{l}l_3l_4;s,0,0} }\int_{\mathbb{C}^{d+2}} D^{d}T_0\,\hA_{l_1l_2l;0,0,s}(T_1;T_2;T_0,D_{R})\hA_{\tilde{l},l_3l_4;s,0,0}(T_0,R;T_3;T_4)\,.
\label{eq:Wls-int-rep}
\end{align}
By using the explicit expression (\ref{eq:SST-3pt}) of the three-point functions, this integral can be rewritten as:
\begin{align}
    \hW_{l,s}^{\{l_i,0\}}&=\frac{1}{\pi^{h}}
\widetilde{\sum_{r,k_{ij}}} T^{\frac{l_1+l_2}{2}-\frac{l+s}{2} +r}_{12} T^{\frac{l_3+l_4}{2}-\frac{l+s}{2} +r}_{34} 
\prod_{(ij)}T_{ij}^{k_{ij}} \int_{\mathbb{C}^{d+2}}
\frac{D^{d}T_0}{T_{01}^{\kappa^1_{l}+r } T_{02}^{\kappa^2_{l}+r }T_{03}^{\kappa^3_{\tilde{l} }+r }T_{04}^{\kappa^4_{\tilde{l}}+r }    }\,,
\label{eq:Wint}
\end{align}
where the summation and the product over the symbol $(ij)$ only take four combinations of indices $i\,,j$, i.~e. $(ij)=(13)\,, (14)\,, (23)\,, (24)$\,.
The summations over $r$ and non-negative integers  $k_{ij}=k_{ji}\,(i,j=1,\dots,4)$ are defined as
\begin{align}
\widetilde{\sum_{r,k_{ij}}} =\frac{s!}{2^s}\sum^{[s/2]}_{r=0} (-1)^{s+r}\frac{(s+h-1)_{-r}}{ r!}
\sum_{\sum_{(ij)}k_{ij}=s-2r}\frac{(-1)^{k_{24}+k_{13}}}{\prod_{(ij)} k_{ij}!}\,.
\label{def:double-sum}
\end{align}
We defined the parameters $\kappa^{1,2}_{l}$ and $\kappa^{3,4}_{\tilde{l}}$ as 
\begin{align}
\begin{split}
\kappa^1_{l}=-\frac{l+s}{2}+\hat{a}+k_{13}+k_{14}\,,\qquad \kappa^2_{l}=-\frac{l+s}{2}-\hat{a}+k_{24}+k_{23}\,,\\
\kappa^3_{\tilde{l}}=-\frac{\tilde{l}+s}{2}-\hat{b}+k_{13}+k_{23}\,,\qquad \kappa^4_{\tilde{l}}=-\frac{\tilde{l}+s}{2}+\hat{b}+k_{14}+k_{24}\,,
\label{def:kappa}
\end{split}
\end{align}
where $\hat{a}=-(l_1-l_2)/2$ and $\hat{b}=(l_3-l_4)/2$\,.
They satisfy the relation:
\begin{align}
\kappa^{1}_{l}+\kappa^{2}_{l}+\kappa^{3}_{\tilde{l}}+\kappa^{4}_{\tilde{l}}+4r=d\,,
\end{align}
which ensures the conformality of the integral.
Therefore, the integral in (\ref{eq:Wint}) may be regarded as the conformal integral except that the integration region is now over a complex null cone. 
In fact, as performed in \cite{Dolan:2011dv}, we can also evaluate it through a Mellin-Barnes transform by assuming
\begin{align}
l\in\mathbb{C}\,, \qquad -d\leq\text{Re}\,l\leq 0\,.  
\end{align}
Here we will skip the details of the calculations, as the steps are somewhat similar to the derivation of AdS Mellin amplitudes (See e.g. \cite{Costa:2012cb,Chen:2017xdz}) and only present the final result which is expressed as a linear combination of the $R$-symmetry block and its shadow, and can be regarded as one of the main results in our work: 
\begin{align}
\hW_{l,s}^{\{l_i,0\}}(T_i)&=T_{12}^{\frac{l_1+l_2}{2}}T_{34}^{\frac{l_3+l_4}{2}}
\left(\frac{T_{24}}{T_{14}}\right)^{\hat{a}}
\left(\frac{T_{13}}{T_{14}}\right)^{\hat{b}}
\left[c_{l,s}\hat{G}_{l,s}^{\hat{a},\hat{b}}(\sigma,\tau)+c_{\tilde{l},s}\hat{G}_{\tilde{l},s}^{\hat{a},\hat{b}}(\sigma,\tau)\right]\,,
\label{eq:R-partial-result1}
\end{align}
where the coefficients $c_{l,s}$ and $c_{\tilde{l},s}$ are given by
{\begin{align}
c_{l,s}&=
\frac{\Gamma(l+h)}{\Gamma(-l)}\frac{\Gamma\left(-\frac{l-s}{2}\pm\hat{b}\right) }{ 
\Gamma\left(-\frac{\tilde{l}-s}{2}\pm\hat{b}\right) } \,,\qquad 
c_{\tilde{l},s}= 
\frac{\Gamma(\tilde{l}+h)}{\Gamma(-\tilde{l})}\frac{\Gamma\left(-\frac{\tilde{l}-s}{2}\pm \hat{a}\right) }{  \Gamma\left(-\frac{l-s}{2}\pm\hat{a}\right)} \,.
\label{eq:c-general}
\end{align}}
Here we also introduced the notation $\Gamma(x\pm y)=\Gamma(x+y)\Gamma(x-y)$\,.
The $R$-symmetry block $\hat{G}_{l,s}^{\hat{a},\hat{b}}(\sigma,\tau)$ and its shadow $\hat{G}_{\tilde{l},s}^{\hat{a},\hat{b}}(\sigma,\tau)$ are:
\begin{align}
\hat{G}_{l,s}^{\hat{a},\hat{b}}(\sigma,\tau)&=
\frac{\Gamma\left(-\frac{\tilde{l}-s}{2}\pm\hat{b}\right) }{ 
\Gamma\left(-\frac{l-s}{2}\pm\hat{b}\right) } \widetilde{\sum_{r,k_{ij}}} \,
\frac{\Gamma(\kappa^3_l+r)\Gamma(\kappa^4_l+r) }{  \Gamma(\kappa^3_{\tilde{l}}+r)\Gamma(\kappa^4_{\tilde{l}}+r)   } 
\sigma^{-\frac{l+s}{2}+r}\tau^{k_{23}}\no\\
&\qquad\times G\left(\kappa^2_{l}+r, \kappa^3_{l}+r;1-l-h ,-l;\sigma,1-\tau\right)\,,
\label{eq:R-block-series}\\
{\hat{G}_{\tilde{l},s}^{\hat{a},\hat{b}}(\sigma,\tau)}&{=
\frac{\Gamma\left(-\frac{l-s}{2}\pm\hat{a}\right) }{ 
\Gamma\left(-\frac{\tilde{l}-s}{2}\pm\hat{a}\right) } \widetilde{\sum_{r,k_{ij}}}\,
\frac{\Gamma(\kappa^1_{\tilde{l}}+r)\Gamma(\kappa^2_{\tilde{l}}+r) }{  \Gamma(\kappa^1_{l}+r)\Gamma(\kappa^2_{l}+r)   } \sigma^{-\frac{\tilde{l}+s}{2}+r}\tau^{k_{23}}}\no\\
&\qquad{\times \,G(\kappa_{\tilde{l}}^2+r,\kappa^3_{\tilde{l}}+r,1-\tilde{l}-h,-\tilde{l};\sigma,1-\tau) \,,}
\label{eq:R-block-series-sh}
\end{align}
where the two variable function $G(\alpha,\beta,\gamma,\delta;x,y)$ is defined as the double power series expansion
\begin{align}
G(\alpha,\beta,\gamma,\delta;x,y)\equiv\sum_{m,n=0}^{\infty}\frac{(\delta-\alpha)_{m}(\delta-\beta)_{m}}{m!(\gamma)_{m}}
\frac{(\alpha)_{m+n}(\beta)_{m+n}}{n!(\delta)_{2m+n}}x^m y^n\,.
\end{align}
This function was originally introduced by Exton in \cite{Exton} and we will call it the Exton function in this paper.
Its properties are summarized in appendix \ref{sec:Extonfun}.
When $s=0$\,, the $R$-symmetry block (\ref{eq:R-block-series}) has a simple form
\begin{align}
\hat{G}_{l,0}^{\hat{a},\hat{b}}(\sigma,\tau)&=\sigma^{-\frac{l}{2}} \,
G\left(-\frac{l}{2}-\hat{a},-\frac{l}{2}-\hat{b};1-l-h ,-l;\sigma,1-\tau\right) \,.
\label{eq:sR-block-G-exp}
\end{align}
We can explicitly check that this function is an eigenfunction of the quadratic Casimir equation for $SO(d+2)$ (see appendix \ref{sec:Extonfun} for details).
In the case of the conformal block, the same expression has been derived in \cite{Dolan:2011dv}.

\paragraph{The $d=4$ $\mathcal{N}=4$ SYM case}
\paragraph{}
Let us consider the $R$-symmetry partial wave (\ref{eq:R-partial-result1}) for four-point correlation function (\ref{eq:4pt-1/2BPS}) of the half-BPS operators in $d=4$ $\mathcal{N}=4$ SYM.

\medskip

As noted in subsection \ref{subsec:setup}, we are interested in the exchanged states that belong to the irreducible representations of $SU(4)_{R}$-symmetry appearing in both operator product expansions of $[0,p_1,0] \otimes [0,p_2,0]$ and $[0,p_3,0] \otimes [0,p_4,0]$\,.
The tensor product of $SU(4)$ representations is decomposed as
\begin{align}
[0,l_1,0]\otimes [0,l_2,0] &=\bigoplus_{m=0}^{\text{min}(l_1,l_2) }\bigoplus_{n=0}^{\text{min}(l_1,l_2)-m}[m, l_2-l_1+2n ,m]\label{eq:SU(4)-rep-split}\,,
\end{align}
where we assumed $l_2\geq l_1$\,.
The right-hand side contains the following exchanged states:
\begin{align}
[0,l_1,0]\otimes [0,l_2,0] \supset\bigoplus_{n=0}^{l_1-s}[s, l_2-l_1+2n ,s]\,.
\end{align}
This indicates that the range of $l-s$ for the exchanged state is given by
\begin{align}
l-s=l_2-l_1+2n \qquad (n=0\,,\dots \,, l_1-s)\,,
\label{eq:l3s3-con}
\end{align}
or equivalently, the combinations of $l_1\,,l_2$\,, and $(l,s)$ satisfies
\begin{align}
\frac{l_1+l_{2}-l-s}{2}\,,\quad \frac{l_2+l+s-l_{1}}{2}\,,\quad \frac{l-s+l_{1}-l_{2}}{2}\in\mathbb{Z}_{\geq0}\,.
\label{eq:selection-three-spin}
\end{align}
 Note that the integrals (\ref{eq:YYY-int}), (\ref{eq:SST-int}) with positive integers $l_i$ would vanish if the quantum numbers $l_i$ do not satisfy the condition (\ref{eq:selection-three-spin}).
Therefore, the $R$-symmetry partial wave (\ref{eq:R-partial-result1}) should be evaluated on the condition
 \begin{align}
\begin{split}
l-s=\begin{cases}
l_2-l_1+2n \qquad &(n=0\,,\dots \,, l_1-s)\\
l_4-l_3+2m \qquad &(m=0\,,\dots \,, l_3-s)
\end{cases}
\,.\label{eq:selection-R-sym}
\end{split}
\end{align}
This rule implies $\hat{a}-\hat{b}\in \mathbb{Z}$\,.
The condition (\ref{eq:selection-R-sym}) is understood as the selection rule of $SO(6)\simeq SU(4)$ representations.

\medskip

In addition to the selection rule (\ref{eq:selection-R-sym}), we have to consider the boundary condition of the $R$-symmetry partial wave (\ref{eq:R-partial-result1}).
Here we further require that the $R$-symmetry partial wave (\ref{eq:R-partial-result1}) has the same boundary condition (\ref{eq:leading-GR}) of the $R$-symmetry block after imposing the selection rule (\ref{eq:selection-R-sym}). 
To this end, we normalize the $R$-symmetry partial wave (\ref{eq:R-partial-result1}) by the factor $1/c_{l,s}$\,.
We next impose the selection rule (\ref{eq:selection-R-sym}) leads to $R$-symmetry partial wave (\ref{eq:R-partial-result1}),
\begin{align}
\frac{1}{c_{l,s}}\hW_{l,s}^{\{l_i,0\}}(T_i)\Bigl\lvert_{(\ref{eq:selection-R-sym})}
&=T_{12}^{\frac{l_1+l_2}{2}}T_{34}^{\frac{l_3+l_4}{2}}
\left(\frac{T_{24}}{T_{14}}\right)^{\hat{a}}
\left(\frac{T_{13}}{T_{14}}\right)^{\hat{b}}\hat{G}_{l,s}^{\hat{a},\hat{b}}(\sigma,\tau)\,,\label{eq:W-normalize}
\end{align}
because the ratio $c_{\tilde{l},s}/c_{l,s}$ vanishes due to the condition (\ref{eq:selection-R-sym}).  
In this sense, the $R$-symmetry partial wave (\ref{eq:four-R-int}) also encodes the $R$-symmetry block.
{It is important to note the similarity between the double trace operator condition $\tau=\Delta-J =\Delta_1+\Delta_2+ 2n\,, n\in\mathbb{Z}_{\geq0}$ and the selection rule (\ref{eq:selection-R-sym}).
An equivalent condition to (\ref{eq:selection-R-sym}) is $-l-s = -(l_1+l_2), \dots ,-(l_1+l_2)+2(l_1-s)$, which is analogous to the double trace operators,  if we identify $-l$ and $s$ with $\Delta$ and $J$,  also $-l_{1,2}$ with $\Delta_{1,2}$ etc.,
but with a cutoff on the descendants due to the compactness of the group $SO(d+2)$. }

\medskip

If we take $s=0$\,, we can show that the $R$-symmetry block (\ref{eq:sR-block-G-exp}) reduces to the known result (\ref{eeq:R-sym-block}).
In order to see the equivalence between this expression and (\ref{eeq:R-sym-block}), we assume $\hat{a}\leq \hat{b}$\,. 
By using the formula (\ref{eq:G-4d})\,, (\ref{eq:sR-block-G-exp}) can be rewritten as
\begin{align}
\hat{G}_{l,0}^{\hat{a},\hat{b}}(\sigma,\tau)
&=\frac{\alpha_1\alpha_2}{\alpha_1-\alpha_2}\left[k^{-\hat{a},-\hat{b}}_{-\frac{l}{2}}(\alpha_1)k^{-\hat{a},-\hat{b}}_{-\frac{l}{2}-1}(\alpha_2)-k^{-\hat{a},-\hat{b}}_{-\frac{l}{2}}(\alpha_2)k^{-\hat{a},-\hat{b}}_{-\frac{l}{2}-1}(\alpha_1) \right]
\,.
\end{align}
If $l$ satisfies the selection rule (\ref{eq:selection-R-sym}), the function $k^{-\hat{a},-\hat{b}}_{-\frac{l}{2}}(\alpha)$ is also expressed in terms of the Jacobi polynomial, 
\begin{align}
   k^{-\hat{a},-\hat{b}}_{-\frac{l}{2}}(\alpha)&=N_{l}^{\hat{a},\hat{b}}\alpha^{\hat{a}}P_{\frac{l}{2}+\hat{a}}^{(\hat{b}-\hat{a},-\hat{a}-\hat{b})}(w)\,,\qquad  \alpha=\frac{2}{1-w}\,,
\end{align}
where the overall constant is given by
\begin{align}
 N_{l}^{\hat{a},\hat{b}}&=(-1)^{\frac{l}{2}+\hat{a}}\frac{\left(\hat{b}-\hat{a}+1\right)_{\frac{l}{2}-\hat{b}}\left(\frac{l}{2}+\hat{a}\right)!}{\left(1+\frac{l}{2}+\hat{b}\right)_{\frac{l}{2}-\hat{b}}(\hat{b}-\hat{a}+1)_{\frac{l}{2}+\hat{a}}}\,.
\end{align}
Therefore, the $R$-symmetry block (\ref{eq:sR-block-G-exp}) becomes
\begin{align}
\hat{G}_{l,0}^{\hat{a},\hat{b}}(\sigma,\tau)
&=2N_{l}^{\hat{a},\hat{b}}N_{l+2}^{\hat{a},\hat{b}}\,(\alpha_1\alpha_2)^{\hat{a}}
P^{(-\hat{a}+\hat{b},-\hat{a}-\hat{b})}_{\frac{l}{2}+\hat{a},\frac{l}{2}+\hat{a}}(\hat{w}_1,\hat{w}_2)
\,,
\end{align}
where $\hat{w}_i\,(i=1,2)$ are defined in (\ref{def:a-w}).
This is the usual expression (\ref{eeq:R-sym-block}) of the $d=4$ $R$-symmetry block \cite{Nirschl:2004pa,Bissi:2015qoa,Doobary:2015gia}. 

\medskip

Our $R$-symmetry block (\ref{eq:R-block-series}) with $s>0$ also reproduces all the known results up to an overall factor,
\begin{align}
    \hat{G}_{l,s}^{(\hat{a},\hat{b})}(\sigma,\tau)\propto (\alpha_1\alpha_2)^{a}\,P_{\frac{l+s}{2}+a,\frac{l-s}{2}+a}^{(-a+b,-a-b)}(\hat{w}_1,\hat{w}_2)\,,
    \label{eq:gJacobi}
\end{align}
where we assumed $\hat{a}\leq \hat{b}$ again.
While we have not been able to show analytically the equivalence between \eqref{eq:R-block-series}, \eqref{eq:R-block-series-sh} and \eqref{eq:gJacobi} for arbitrary $l$ and $s$,
we have explicitly checked their equivalence in appendix \ref{sec:Rblock} 
for $s\leq l\leq 4\,, 0\leq s\leq2$\,, up to the overall factors. This explicit check is somewhat non-trivial as we also need to perform the summation over the four fold partition of integers for given $l$ ({For the explicit expressions, see (\ref{eq:s=0-1})-(\ref{eq:s=2-5}})).
Note that $\hat{a}, \hat{b}$ are taken to satisfy the selection rule (\ref{eq:selection-R-sym}).

\paragraph{The $d=3$ case}
\paragraph{}

Finally, let us comment on the connection of our result with superconformal field theories in other dimensions.
We focus here on a six dimensional $(2, 0)$ superconformal theory which is holographically dual to eleven dimensional supergravity on $AdS_{7}\times S^4$\,.
The associated superconformal algebra is $\mathfrak{osp}(8^*
|4)$ that contains a bosonic subgroup $\mathfrak{usp}(4) \simeq\mathfrak{so}(5)$ as the $R$-symmetry.

\medskip

The four-point functions of the half-BPS operators, that transform in the $[2,0]$ representation of $\mathfrak{so}(5)$\,, and its OPE structures have been considered by many authors \cite{Ferrara:2001uj,Dolan:2004mu,Beem:2015aoa,Heslop:2004du,Rastelli:2017ymc}.
As in the previous case, the four-point functions are expanded in terms of the $\mathfrak{so}(5)$ spherical harmonic functions.
For a consistency check, we confirmed that our $R$-symmetry block (\ref{eq:R-block-series}) reproduces the known results listed in (B.14) of \cite{Nirschl:2004pa} up to overall constants:
\begin{align}
\begin{split}
\hat{G}_{1,1}^{0,0}(\sigma,\tau)&=-\frac{2}{3}\left(\frac{1}{\sigma}-\frac{\tau}{\sigma}\right)
    =-\frac{2}{3}Y_{1\,0}\left(\frac{1}{\sigma},\frac{\tau}{\sigma}\right)\,,\\
    \hat{G}_{2,0}^{0,0}(\sigma,\tau)&=\frac{1}{2}\left(\frac{1}{\sigma}+\frac{\tau}{\sigma}-\frac{2}{5}\right)
    =\frac{1}{2}Y_{1\,1}\left(\frac{1}{\sigma},\frac{\tau}{\sigma}\right)\,,\\
    \hat{G}_{2,2}^{0,0}(\sigma,\tau)&=\frac{32}{45}\left(\frac{1}{\sigma^2}+\left(\frac{\tau}{\sigma}\right)^2-\frac{2\tau}{\sigma^2}-\frac{2}{3}\left(\frac{\tau}{\sigma}+\frac{1}{\sigma}\right)+\frac{1}{6}\right)
    =\frac{32}{45}Y_{2\,0}\left(\frac{1}{\sigma},\frac{\tau}{\sigma}\right)\,,\\
    \hat{G}_{3,1}^{0,0}(\sigma,\tau)&=-\frac{28}{105}\left(\frac{1}{\sigma^2}-\left(\frac{\tau}{\sigma}\right)^2-\frac{4}{7}\left(\frac{1}{\sigma} -\frac{\tau}{\sigma}\right)\right)
    =-\frac{28}{105}Y_{2\,1}\left(\frac{1}{\sigma},\frac{\tau}{\sigma}\right)\,,\\
    \hat{G}_{4,0}^{0,0}(\sigma,\tau)&=\frac{1}{6}
    \left(\frac{1}{\sigma^2}+\left(\frac{\tau}{\sigma}\right)^2+\frac{4\tau}{\sigma^2}-\frac{8(1+\tau)}{9\sigma}+\frac{8}{63}\right)
    =\frac{1}{6}Y_{2\,2}\left(\frac{1}{\sigma},\frac{\tau}{\sigma}\right)\,.
    \end{split}
\end{align}
The $R$-symmetry block $\hat{G}_{l,s}^{0,0}(\sigma,\tau)$ corresponds to the $\mathfrak{so}(5)$ spherical harmonic function $Y_{\frac{l+s}{2}\,\frac{l-s}{2}}\left(1/\sigma,\tau/\sigma\right)$  associated with the $\mathfrak{usp}(4)$ Dynkin labels $[l-s, 2s]$\,.
In this way, the $R$-symmetry block (\ref{eq:R-block-series}) works well for the odd $d$ dimension cases.

\section{Superconformal Casimir equation and the Heckman-Opdam systems}\label{sec:HO}
\paragraph{}
In this section, we will establish the relation between the quadratic superconformal Casimir equation (\ref{eq:long-Casimireq}) and the Heckman-Opdam (HO) hypergeometric systems associated with the BC${}_2$ root system. This is a generalization of the similar relation for the conformal Casimir equation discovered in \cite{Isachenkov:2016gim},
\cite{Isachenkov:2017qgn},
\cite{Chen:2016bxc}.

\medskip

The HO hypergeometric function associated with the BC${}_2$ root system is a solution to the partial differential equation (for more details, see \cite{HO1,Hec1,O1,O2,Isachenkov:2016gim,Isachenkov:2017qgn}),
\begin{align}
L_{\rm BC_2}(k)\,\varphi(k;w_i)=\left(\langle  \la,\la \rangle-\langle  \rho(k),\rho(k) \rangle\right)\,\varphi(k;w_i)\,,
\label{eq:HOeq-sec5}
\end{align}
where the differential operator $L_{\rm BC_2}(k)$ is
\begin{align}
L_{\rm BC_2}(k)&=-\sum_{i=1}^2\left[(1-w^2_i)\frac{\partial^2}{\partial w^2_i}-(k_s+(k_s+2k_l+1)w_i)\frac{\partial}{\partial w_i}\right]\no\\
&\quad- \frac{2k_{m}}{w_1-w_2}
\left((1-w^2_1)\frac{\partial}{\partial w_1}-(1-w^2_2)\frac{\partial}{\partial w_2}\right)\,.
\label{eq:HOop-sec5}
\end{align}
The equation (\ref{eq:HOeq-sec5}) is characterized by three complex parameters $k_{s}\,, k_{m}\,,$ and $k_{l}$ which are associated with the short, middle, and long positive roots of BC${}_2$ root system, respectively.
The bracket $\langle \cdot, \cdot \rangle$ is the inner product in the two-dimensional Euclidean space, and $\la$ and $\rho_{\rm BC_2}(k)$ are defined as
\begin{align}
    \la&=\la_1\,e_1+\la_2\,e_2\,,\\
\rho_{\rm BC_2}(k)
&=\left(\rho_{\rm BC_1}(k)+k_{m}\right)e_1+\rho_{\rm BC_1}(k)\,e_2\,,\qquad \rho_{\rm BC_1}(k)= \frac{1}{2}(k_{s}+2k_{l})\,,
\end{align}
where $e_{i}$ are orthonormal vectors satisfying $\langle e_i, e_j \rangle=\delta_{ij}$\,.

\medskip

As observed in \cite{Isachenkov:2016gim,Isachenkov:2017qgn}, any $d$-dimensional scalar conformal block can be described in the HO system (\ref{eq:HOeq-sec5}) associated with the BC${}_2$ root system.
This observation can easily be extended to the case of the superconformal block for the long multiplets.
In the following discussions, we will explain it.

\subsection{Superconformal blocks as HO hypergeometric function}
\paragraph{}
We will consider the reduced part of the superconformal blocks $H^{a,b}_{\Delta,J,l,s}(z_i,\alpha_i)$ for the long multiplets with the $SU(4)$ Dynkin index $[s,l-s,s]$ satisfying (\ref{eq:long-Casimireq}).

\medskip

The differential operators $\check{\cD}_{4}^{(a,b)}$ and $\hat{\cD}_{4}^{(a,b)}$ defined in (\ref{defchD}) and (\ref{defhatD}) are  related to the differential operator $L_{\rm BC_2}$ by performing the following similarity transformations\footnote{In this case, we take $\check{a}=a\,, \check{b}=b$ and $\hat{a}=a\,, \hat{b}=b$\,.},
\begin{align}
(z_1z_2)^{a} \check{\cD}_{4}^{(a,b)}(z_1z_2)^{-a}&=\check{L}_{\rm BC_2}(\check{k})
+a(2a+4)\,,\\
(\alpha_1\alpha_2)^{-a}\hat{\cD}_{4}^{(a,b)} (\alpha_1\alpha_2)^{a}&=\hat{L}_{\rm BC_2}(\hat{k})
+a(2a-4)\,,
\label{eq:similarity-tr-super}
\end{align}
where variables $z_i$ and $\alpha_i$ are related to $\check{w}_i$ and $\hat{w}_i$ in $\check{L}_{\rm BC_2}$ and $\hat{L}_{\rm BC_2}$ by
\begin{align}
z_i=\frac{2}{1-\check{w}_i}\,, \qquad \alpha_i=\frac{2}{1-\hat{w}_i}\,.
\label{eq:za-w}
\end{align}
In these HO systems, the each multiplicity functions $\check{k}\,, \hat{k}$ take values
\begin{align}
\check{k}_s&=-2b\,,\qquad \check{k}_l=a+b+\frac{1}{2}\,, \qquad \check{k}_{m}=1\,,\\
\hat{k}_s&=2b\,,\qquad \hat{k}_l=-a-b+\frac{1}{2}\,, \qquad \hat{k}_{m}=1\,.
\end{align}
From the similarity transformations (\ref{eq:similarity-tr-super}), let us define
\begin{align}
\check{G}_{\Delta+4,J}^{a,b}(z_i)&\equiv (z_1z_2)^{-a} \check{\varphi}(\check{w}_i)\,,\qquad
\hat{G}^{a,b}_{l,s}(\alpha_i)\equiv(\alpha_1\alpha_2)^{a}\hat{\varphi}(\hat{w}_i)\,,
\label{eq:block-to-HO}
\end{align}
or equivalently, 
\begin{align}
\begin{split}
H^{a,b}_{\Delta,J,l,s}(z_i,\alpha_i)&\equiv (z_1z_2)^{-a} (\alpha_1\alpha_2)^{a}  \check{\varphi}(\check{w}_i)\hat{\varphi}(\hat{w}_i)\,.
\end{split}
\end{align}
Then, the quadratic superconformal Casimir equation (\ref{eq:long-Casimireq}) can be rewritten as
\begin{align}
\check{L}_{\rm BC_2}(\check{k})\check{\varphi}(\check{w}_i)
&=\left[\frac{1}{2}\left(\Delta(\Delta+4)+J(J+2)\right)-a(2a+4)\right]\check{\varphi}(\check{w}_i)\,,\label{eq:sconf-HO}\\
\hat{L}_{\rm BC_2}(\hat{k})\hat{\varphi}(\hat{w}_i)
&=\left[\frac{1}{2}\left(l(l+4)+s(s+2)\right)-a(2a-4)\right]\hat{\varphi}(\hat{w}_i)\,.\label{eq:sR-HO}
\end{align}
By comparing the HO systems  (\ref{eq:HOeq-sec5}) with the above equations, the spectral parameters $\check{\la}\,, \hat{\la}$ for each system are taken as
\begin{align}
\check{\la}_1&=-\frac{\Delta-J+1}{2}\,,\qquad 
\check{\la}_2=-\frac{\Delta+J+3}{2}\,,\label{eq:la-DJ-id}\\
\hat{\la}_1&=\frac{l+s+3}{2}\,,\qquad 
\hat{\la}_2=\frac{l-s+1}{2}\,.\label{eq:la-S-id}
\end{align}
In this way, the superconformal Casimir equation (\ref{eq:long-Casimireq}) can be translated to two copies of BC${}_2$ HO hypergeometric equations.

\medskip

Finally, let us comment on the relation between HO system and Calogero-Sutherland (CS) model.
It is well known that a given HO system is related to a CS model associated with the same root system by a similarity transformation \cite{HO1}. 
We can show that the eigenvalues of the CS Hamiltonians corresponding to $\check{L}_{\rm BC_2}$ and $\hat{L}_{\rm BC_2}$ are given by
\begin{align}
E_{\rm CS}(\check{\la})&=-\sum_{i=1}^2 (\check{\la}_i)^2
=-\frac{1}{2}\left(C_{\Delta+4,J}+5\right)\,,\\
E_{\rm CS}(\hat{\la})&=-\sum_{i=1}^2 (\hat{\la}_i)^2
=-\frac{1}{2}\left(C_{l,s}+5\right)\,.\label{eq:CS-eigenS}
\end{align}
Since the eigenvalue of the conformal Casimir operator takes continuous values, the conformal case corresponds to the scattering problem of the CS model.
On the other hand, the $R$-symmetry case describes a bound state problem because its eigenvalue (\ref{eq:CS-eigenS}) takes discrete values.

\subsection{$R$-symmetry part}
\paragraph{}
The relation between the conformal block and the HO hypergeometric system has been discussed in \cite{Isachenkov:2017qgn}, so we will skip it and focus on a solution of the HO hypergeometric system (\ref{eq:sR-HO}) for the $R$-symmetry part.
As we will show, the generalized Jacobi polynomial (\ref{def:Jacobi-4d}) or equivalently the $R$-symmetry block  can also be described in the HO hypergeometric system associated with the BC${}_2$ root system.

\medskip

The solutions of the HO hypergeometric system are constructed from the liner combinations of the Harish-Chandra series $\Phi_{\rm BC_2}(\hat{\la},\hat{k};\hat{w})$ and its BC${}_2$ Weyl transformed ones \cite{HO1} (See also \cite{Isachenkov:2017qgn}).
The BC${}_2$ Weyl group is generated by
\begin{align}
\hat{\omega}_1:(\hat{\la}_1,\hat{\la}_2)\to (\hat{\la}_2,\hat{\la}_1)\,,\qquad 
\hat{\omega}_2:(\hat{\la}_1,\hat{\la}_2)\to (\hat{\la}_1,-\hat{\la}_2)\,,
\label{eq:sphere-BC2-weyl}
\end{align}
or equivalently,
\begin{align}
\hat{\omega}_1:(l,s)\to (l,-2-s)\,,\qquad 
\hat{\omega}_2:(l,s)\to (-1+s,1+l)\,,
\end{align}
where we used the relation (\ref{eq:sR-HO})\footnote{$\hat{\omega}_2\,\hat{\omega}_1\,\hat{\omega}_2$ gives the shadow transformation of $l$ as introduced in subsection \ref{subsec:split}}.

\medskip

As in the conformal case, we consider a linear combination of $\Phi_{\rm BC_2}(\hat{\la},\hat{k};\hat{w})$ and the Weyl transformed HO series $\Phi_{\rm BC_2}(\hat{\la},\hat{k};\hat{\omega}_1\hat{w})$ with $\hat{k}_{m}=1$,
\begin{align}
\hat{\varphi}(\hat{w}_i)&=c_{\rm BC_2}(\hat{\la},\hat{k})
\Phi_{\rm BC_2}(\hat{\la},\hat{k};\hat{w})+c_{\rm BC_2}(\hat{\omega}_1\hat{\la},\hat{k})
\Phi_{\rm BC_2}(\hat{\omega}_1\hat{\la},\hat{k};\hat{w})\no\\
&=\frac{4(\hat{k}_s+\hat{k}_l+\frac{1}{2})}{\hat{\la}_1{}^2-\hat{\la}_2{}^2}
\left(\prod^2_{j=1}c_{\rm BC_1}(\hat{\la}_{j},\hat{k})\right)\no\\
&\quad \times
\frac{\Phi_{\rm BC_1}(\hat{\la}_1,\hat{k};\hat{w}_1)\Phi_{\rm BC_1}(\hat{\la}_2,\hat{k};\hat{w}_2)
-\Phi_{\rm BC_1}(\hat{\la}_1,\hat{k};\hat{w}_2)\Phi_{\rm BC_1}(\hat{\la}_2,\hat{k};\hat{w}_1)}{\hat{w}_1-\hat{w}_2}\no\\
&\propto \frac{(\alpha_1\alpha_2)^{1-a}}{\alpha_1-\alpha_2}\left[k^{-a,-b}_{-\frac{l-s}{2}}(\alpha_1)k^{-a,-b}_{-\frac{l+s}{2}-1}(\alpha_2)-k^{-a,-b}_{-\frac{l-s}{2}}(\alpha_2)k^{-a,-b}_{-\frac{l+s}{2}-1}(\alpha_1) \right]
\,.
\label{eq:BC-HO-sh2}
\end{align}
Since the BC${}_1$ Harish-Chandra series $\Phi_{\rm BC_1}(\hat{\la}_i,\hat{k};\hat{w}_i)$ is proportional to the Jacobi polynomial,
\begin{align}
c_{\rm BC_1}(\hat{\la}_{1},\hat{k})\Phi_{\rm BC_1}(\hat{\la}_1,\hat{k};\hat{w}_1)&\propto
\alpha_1^{-a}P_{\frac{l+s}{2}+a+1}^{(-a+b,-a-b)}(\hat{w}_1)\,,\label{eq:BC1-Jacboi-1}\\
c_{\rm BC_1}(\hat{\la}_{2},\hat{k})\Phi_{\rm BC_1}(\hat{\la}_2,\hat{k};\hat{w}_2)&\propto
\alpha_2^{-a}P_{\frac{l-s}{2}+a}^{(-a+b,-a-b)}(\hat{w}_2)\label{eq:BC1-Jacboi-2}
\,,
\end{align}
the HO hypergeometric function (\ref{eq:BC-HO-sh2}) is shown to be the generalized Jacobi polynomial 
\begin{align}
\hat{\varphi}(\hat{w}_i)&\propto
(\alpha_1\alpha_2)^{-a}P^{(-a+b,-a-b)}_{\frac{l+s}{2}+a,\frac{l-s}{2}+a}(\hat{w}_i)\,,
\label{eq:BC_2-Jacobi}
\end{align}
after requiring that $l$ and $s$ satisfy the selection rule (\ref{eq:selection-R-sym}).
In this way, the $R$-symmetry blocks (\ref{eeq:R-sym-block}) for the long multiplets can be described in the HO hypergeometric system associated with the BC${}_2$ root system.


\acknowledgments
We would like to thank P.\,Heslop and T.\,Nishioka for useful discussions, and Kyoto university high energy theory group and university of Tokyo high energy theory group for the hospitality while this work was being completed.
The work of H.\,Y.\,C. was supported in part by Ministry of Science and Technology (MOST) through the grant
108-2112-M-002 -004 -.
The work of J.\,S. was supported in part by Ministry of Science and Technology (project no. 108-2811-M-002-528), National Taiwan University, and  Osaka City University Advanced Mathematical Institute (MEXT Joint Usage/Research Center on Mathematics and Theoretical Physics).

\appendix

\section*{The Appendices}

\section{The details of embedding formalism of $S^{d+1}$}\label{App:embedding}
\paragraph{}
In this appendix, we will present more details of the embedding formalism on $S^{d+1}$ used in the main text.

\subsection{Bulk side}
\paragraph{}
Let us first consider a symmetric traceless rank-$s$ tensor field on $\mathbb{R}^{d+2}$ with the components $h_{\hA_1\cdots \hA_{s}}(\hX)$\,.
The tensor field can be restricted to $S^{d+1}$ by using the projection operator:
\begin{align}
(\cP h)_{\hA_1\cdots \hA_{s}}(\hX)\equiv\hat{G}_{\{\hA_1}{}^{\hB_1}\cdots \hat{G}_{\hA_s\}}{}^{\hB_s}h_{\hB_1\cdots \hB_{s}}(\hX)\,,
\label{def:proj-to-S}
\end{align}
where $\hat{G}_{\hA\hB}$ is the induced metric
\begin{align}
\hat{G}_{\hA\hB}=\delta_{\hA\hB}-\hX_{\hA}\hX_{\hB}\,.
\end{align}
Such that a null vector with respect to the induced metric:
\begin{equation}
\hat{G}_{\hA\hB}\hX^{\hA} \hX^{\hB} = 0 \leftrightarrow \hX^2 =1 
\end{equation}
describes a unit sphere.
In fact, the projection operator guarantees that $(\cP h)_{\hA_1\cdots \hA_{s}}(\hX)$ is transverse to the hypersurface $\hX^2=1$ i.e. 
\begin{align}
X^{\hA_1}(\cP h)_{\hA_1\cdots \hA_{s}}(\hX)=0\,.
\label{eq:trans-cond}
\end{align}
In other words, $(\cP h)_{\hA_1\cdots \hA_{s}}(\hX)$ is the spherical analogue of so-called symmetric, traceless and transverse (STT) tensor.
This condition also implies the tensor field $h_{\hA_1\cdots \hA_{s}}(\hX)$ has a gauge symmetry
\begin{align}
h_{\hA_1\cdots \hA_{s}}(\hX)\sim h_{\hA_1\cdots \hA_{s}}(\hX)+\hX_{\{\hA_1}\Psi_{\hA_2\dots \hA_s\}}(\hX)\,,
\label{eq:trans-cond-gauge}
\end{align} 
where $\Psi_{\hA_1\dots \hA_{s-1}}(\hX)$ is any rank-$(s-1)$ tensor field on $S^{d+1}$ and $\{\dots\}$ denotes traceless symmetrization of indices, as they project to the same symmetric traceless tensor on $S^{d+1}$.
In this sense, the term containing sub-leading $\Psi_{\hA_1\dots \hA_{s-1}}(\hX)$ is regarded as an un-physical mode of $h_{\hA_1\cdots \hA_{s}}(\hX)$\,.

\medskip

In the direct analogy with the AdS space tensor, there is an efficient way to implement index contraction on tensor fields on $S^{d+1}$\,.
That is to express such a field as a polynomial of the auxiliary null vector $U^{\hA}$:
\begin{align}
h_{s}(\hX, U)\equiv\frac{1}{s!}h_{\hA_1\cdots \hA_{s}}(\hX)U^{\hA_1}\cdots U^{\hA_s}\,,\qquad U\cdot U\equiv \sum_{\hA,\hB=1}^{d+2}\delta_{\hA\hB}\,U^{\hA}U^{\hB}=0\,.
\end{align}
Since $S^{d+1}$ is embedded in the Euclidean space $\mathbb{R}^{d+2}$, the null vector $U^{\hA}$ should be complex i.e. $U^{\hA}\in \mathbb{C}^{d+2}$\,.
Furthermore, in order to restrict the tensor field on $S^{d+1}$\,, we assume
\begin{align}
U\cdot \hX=\sum_{\hA,\hB=1}^{d+2}\delta_{\hA\hB}\,U^{\hA}X^{\hB}=0\,.
\end{align}
The rank of the tensor field $h_{\hA_1\cdots \hA_{s}}(\hX)$ is then translated into the homogeneity of the polynomial $h_{s}(\hX, U)$, as encoded in the equation:
\begin{align}
(U\cdot \partial_{U}-s)h_{s}(\hat{X},U)=0\,.
\end{align}
The (projected) components $(\mathcal{P}h)_{\hA_1\cdots \hA_{s}}(\hX)$ can be reproduced by using the differential operator:
\begin{align}
\hat{K}_{\hA}&\equiv\left(\frac{d-1}{2}+U\cdot \frac{\partial}{\partial U}\right)\left(\frac{\partial}{\partial U^{\hA}}-\hX_{\hA}\left(\hX\cdot\frac{\partial}{\partial U}\right) \right)\no\\
&\quad 
-\frac{1}{2}U_{\hA}\left(\frac{\partial^2}{\partial U \cdot \partial U}-\left(\hX\cdot \frac{\partial}{\partial U}\right)\left(\hX\cdot \frac{\partial}{\partial U}\right)\right)\,.
\label{def:K-proj}
\end{align}
This operator is constructed by requiring 
\begin{align}
\text{transverse:}\quad \hX^{\hA}\hat{K}_{\hA}=0\,,\quad
\text{traceless:}\quad \hat{K}^{\hA}\hat{K}_{\hA}=0\,,\quad
\text{symmetric:}\quad [\hat{K}_{\hA},\hat{K}_{\hB}]=0\,.
\end{align}
We can check that the above differential operator (\ref{def:K-proj}) satisfies
\begin{align}
\hat{G}_{\{\hA_1}{}^{\hB_1}\cdots \hat{G}_{\hA_s\}}{}^{\hB_s}=\frac{1}{s!\left(\frac{d-1}{2}\right)_s}\hat{K}_{\hA_1}\cdots \hat{K}_{\hA_s}U^{\hB_1}\cdots U^{\hB_s}\,,
\label{eq:proj-s-sym}
\end{align}
where $(x)_{n}$ is defined as $(x)_{n}=\frac{\Ga(x+n)}{\Ga(x)}$\,.
Therefore, we obtain
\begin{align}
(\cP h)_{\hA_1\cdots \hA_{s}}(\hX)=\frac{1}{s!\left(\frac{d-1}{2}\right)_s}\,\hat{K}_{\hA_1}\cdots \hat{K}_{\hA_s}h_{s}(\hX, U)\,.
\end{align}
Note that if the operator $\hat{K}_{\hA}$ acts on such a polynomial $h_{s}(\hX, U)$, this operator is effectively simplified to
\begin{align}
\hat{K}_{\hA}=\left(\frac{d-1}{2}+U\cdot \frac{\partial}{\partial U}\right)\frac{\partial}{\partial U^{\hA}}\,.
\end{align}

\medskip

Finally, let us introduce the covariant derivative on $S^{d+1}$ defined by
\begin{align}
\hat{\nabla}_{\hA}\equiv \hat{G}_{\hA}{}^{\hB}\frac{\partial }{\partial \hX^{\hB}}
-\frac{\hX^{\hB}}{\hX^2} \hat{\Sigma}_{\hA\hB}\,,\qquad 
\hat{\Sigma}_{\hA\hB}=U_{\hA}\frac{\partial}{\partial U^{\hB}}-U_{\hB}\frac{\partial}{\partial U^{\hA}}\,.
\label{def:covariant-der}
\end{align}
Here $\hat{\Sigma}_{\hA\hB}$ are Lorentz generators in the differential representation acting on null vectors $U^{\hat{A}}$\,.
The covariant derivative acting on the projected tensor field $(\cP h)_{\hA_1\cdots \hA_{s}}(\hX)$ can be written as
\begin{align}
\hat{\nabla}_{\hB}(\cP h)_{\hA_1\cdots \hA_{s}}(\hX)
&=\hat{G}_{\hB}{}^{\hC}\hat{G}_{\{\hA_1}{}^{\hC_1}\cdots \hat{G}_{\hA_s\}}{}^{\hC_s}\frac{\partial }{\partial \hX^{\hB}}h_{\hC_1\cdots \hC_{s}}(\hX)\,,
\end{align}
where we used the transversality (\ref{eq:trans-cond}) of $(\cP h)_{\hA_1\cdots \hA_{s}}(\hX)$\,.

\subsection{``Boundary'' side}
\paragraph{}
Next, we will discuss the embedding formalism for the ``boundary" side of the sphere, which is described by the complex null cone (\ref{def:c-cone}) with complex coordinates $\{T_{\hA}\}$.

\medskip

Let $F_{\hA_1\dots \hA_s}(T)$ be an arbitrary tensor field on the complex plane $\mathbb{C}^{d+2}$.
The physical tensor fields on  the complex null cone surface (\ref{def:c-cone}) can be obtained by restricting the tensor fields $F_{\hA_1\dots \hA_s}(T)$ on $\mathbb{C}^{d+2}$ to the complex null cone surface (\ref{def:c-cone}).
This is performed by using the projection operator
\begin{align}
\hat{\Pi}_{\hA_1 \dots \hA_s}{}^{\hB_1\dots \hB_s}&\equiv \hat{\Pi}_{i_1 \dots i_l}{}^{j_1\dots j_s}\frac{\partial T_{\hA_1}}{ \partial y_{i_1}}\cdots \frac{\partial T_{\hA_s}}{ \partial y_{i_l}}
\frac{\partial T^{\hB_1}}{ \partial y^{j_1}}\cdots \frac{\partial T^{\hB_s}}{ \partial y^{j_s}}\,,\\
\hat{\Pi}_{i_1 \dots i_s}{}^{j_1\dots j_s}&\equiv\delta^{j_1}_{\{i_1}\cdots \delta_{i_s\}}^{j_s}\,.
\end{align}
In fact, from the fact $T^{\hA}\frac{\partial T_{\hA}}{ \partial y_{i}}=0$, the projected tensor field satisfies the transversality condition
\begin{align}
T^{\hA_1}(\hat{\Pi} F)_{\hA_1\dots \hA_s}(T)=0\,.
\end{align}
Here for convenience, we introduced a parameterization of the null vector
\begin{align}
T^{\hA}=\left(T^{1},T^{2},T^{k}\right)=\left(\frac{i}{2}\left(1+\overrightarrow{y}^2\right),\frac{1}{2}\left(1-\overrightarrow{y}^2\right), y^k \right)\,,
\label{eq:T-y}
\end{align}
where $\overrightarrow{y}\in \mathbb{R}^{d}\,, \overrightarrow{y}^2\equiv \sum_{k=1}^{d}(y^k)^2$\,.
This parameterization is regarded as the counterpart of the Poincar\'e boundary coordinates for  $AdS_{d+1}$, but notice that we now have complex entries instead.
The tensor fields on the complex null cone can be obtained by using the projection:
\begin{align}
F_{j_1\dots j_s}(y)\equiv\frac{\partial T^{\hA_1}}{ \partial y^{j_1}}\cdots \frac{\partial T^{\hA_s}}{ \partial y^{j_s}}F_{\hA_1\dots \hA_s}(T)\,,\qquad 
\frac{\partial T^{\hA}}{ \partial y^{j}}&=\left(~i y_{j},\,- y_{j},\,\delta^i_{j}~\right)\,.
\end{align}
In particular, the induced metric is
\begin{align}
\delta_{jk}^{(\mathbb{C} )}=\frac{\partial T^{\hA}}{ \partial y^{j}}\frac{\partial T^{\hB}}{ \partial y^{k}}\delta_{\hA\hB}\,.
\end{align}

\medskip

The tensor fields on the embedding space can also be described as a homogeneous polynomial of the null vector $R^{\hA}$ satisfying $T\cdot R=0$,
\begin{align}
F_{s}(T,R)=\frac{1}{s!}F_{\hA_1\dots \hA_s}(T)R^{\hA_1}\cdots R^{\hA_s}\,.
\end{align}
As in the previous subsection, we can construct a differential operator similar to $\hat{K}_{\hA}$ defined in (\ref{def:K-proj})
\begin{align}
\hat{D}_{R,\hA}&=\left(\frac{d-2}{2}+R\cdot\frac{\partial}{\partial R} \right)\frac{\partial}{\partial R^{\hA}}
-\frac{1}{2}R_{\hA}\frac{\partial^2}{\partial R \cdot \partial R}\,,
\label{def:DR}
\end{align}
which satisfies 
\begin{align}
&\hat{D}_{R}\cdot\hat{D}_{R}=0\,,\qquad
 [\hat{D}_{R,\hA}, \hat{D}_{R,\hB}]=0\,,\\
&\hat{\Pi}_{\hA_1 \dots \hA_s}{}^{\hB_1\dots \hB_s} =\frac{1}{s!\left(\frac{d-2}{2}\right)_s}\hat{D}_{R,\hA_1}\cdots \hat{D}_{R,\hA_l}R^{\hB_1}\cdots R^{\hB_s}\,.
\label{eq:proj-b-sym}
\end{align}
Therefore, the (projected) components $(\hat{\Pi}F_{s})_{\hA_1\dots \hA_s}$ can be reproduced from $F_{s}(T,R)$
\begin{align}
(\hat{\Pi}F_{s})_{\hA_1\dots \hA_s}=\frac{1}{s!\left(\frac{d-2}{2}\right)_s}\hat{D}_{R,\hA_1}\cdots \hat{D}_{R,\hA_s}F_{s}(T,R)\,.
\end{align}

\section{Spherical harmonic functions}\label{app:SHfun}
\paragraph{}
In this appendix, we will give some properties of the scalar and tensor spherical harmonic functions.

\subsection{Scalar case}
\paragraph{}
Let us first see the scalar spherical harmonic (SSH) function $K_{l,0}^{\sfs}(\hX;T)$ given by (\ref{def:shf}).
By using the expression of the covariant derivative (\ref{def:covariant-der}), we can check that (\ref{def:shf}) is a solution to the equation
\begin{align}
\hnabla^2K_{l,0}^{\sfs}(\hX;T)=-l(l+d)K_{l,0}^{\sfs}(\hX;T)\,,
\end{align}
or equivalently satisfies the quadratic Casimir equation for $SO(d+2)$ with the eigenvalue $l(l+d)$\,.
In the following discussion, we will show the orthogonality relation and three-point functions involving only the SSH functions.

\subsubsection*{Orthogonality relation}
\paragraph{}
The SSH function (\ref{def:shf}) satisfies the orthogonality relation,
\begin{align}
I_{ll'}(T_1,T_2)\equiv\int_{S^{d+1}}\dd^{d+2}{\bf \hX}\,K_{l,0}^{\sfs}(\hX;T_1)K_{l',0}^{\sfs}(\hX;T_2)
=Z_{l}\,\hC_{l,0}^2\delta_{l,l'}\,T_{12}^{l}\,,
\label{eq:orthogonal-S}
\end{align}
where the overall constant $Z_{l}$ is given by
\begin{align}
Z_{l}=(-1)^l\frac{l!\,\text{Vol}(S^{d+1})}{2^{2l}(h+1)_{l}}\,,\qquad h= \frac{d}{2}\,.
\end{align}
The integration measure is symmetric under $SO(d+2)$ transformations and defined as
\begin{align}
\int_{S^{d+1}}\dd^{d+2}{\bf \hX}=\int_{S^{d+1}}\dd^{d+2}\hX\,\delta(\hX^2-1)\,.\label{eq:measure-S}
\end{align}
The relation (\ref{eq:orthogonal-S}) is regarded as the spherical counterpart of joining two bulk to boundary AdS propagators to obtain the two-point CFT correlation functions.

\medskip

The orthogonality of the SSH functions easily follows from the $SO(d+2)$ symmetry.
In fact, if we let $g$ be an element of $SO(d+2)$\,, $I_{ll'}(T_1,T_2)$ transform as
\begin{align}
    I_{ll'}(g\cdot T_1,g\cdot T_2)=I_{ll'}(T_1,T_2)\,.
\end{align}
This implies $I_{ll'}(T_1,T_2)$ only depends on $T_{12}$ and vanishes if $l\neq l'$\,.
Next, let us compute the overall constant $Z_{l}$. 
To this end, we will consider the following integration:
\begin{align}
I^{\hA_1\dots \hA_{2l}}\equiv\int_{S^{d+1}}\dd^{d+2}{\bf \hX}\,\hX^{\hA_1}\cdots\hX^{\hA_{2l}}\,.
\end{align}
By using the $SO(d+2)$ symmetry, this integration should have the form
\begin{align}
I^{\hA_1\dots \hA_{2l}}=N_{l}\, 
\delta^{(\hA_1\hA_2}\cdots \delta^{\hA_{2l-1}\hA_{2l}) }\,,
\label{eq:Int-I-1}
\end{align}
where the constant $N_{l}$ is determined from a contraction:
\begin{align}
\text{Vol}(S^{d+1})=\delta_{\hA_1\hA_{2}}\cdots \delta_{\hA_{2l-1}\hA_{2l} }I^{\hA_1\dots \hA_{2l}}
=N_{l}\frac{2^{2l}l!}{(2l)!}\left(h+1\right)_{l}\,.
\end{align}
From this result, the overall constant $Z_{l}$ can be evaluated as
\begin{align}
I_{ll}&=\,\hC_{l,0}^2\,T_{1}^{\hA_1}\cdots T_{1}^{\hA_{l}} T_{2}^{\hA_{l+1}}\cdots T_{2}^{\hA_{2l}} I^{\hA_1\dots \hA_{2l}} =N_{l}\,\hC_{l,0}^2(-1)^{l}\frac{(l!)^2}{(2l)!}\,T_{12}^{l},\no\\
&=\hC_{l,0}^2(-1)^l\frac{l!\,\text{Vol}(S^{d+1})}{2^{2l}\left(h+1\right)_{l}}T_{12}^{l}\,.
\end{align}

\subsubsection*{Three-point functions}
\paragraph{}
Next let us evaluate the spherical scalar three-point function given by the following integral:
\begin{align}
\hA_{l_1l_2l_3}(T_i)=\int_{S^{d+1}}\dd^{d+2}{\bf \hX}\,
 K_{l_1,0}^{\sfs}(\hX;T_1)K_{l_2,0}^{\sfs}(\hX;T_2)K_{l_3,0}^{\sfs}(\hX;T_3)\,.
\label{eq:three-point-KKK}
\end{align}
This integral can be evaluated by using the formula (\ref{eq:Int-I-1}).
Instead of doing this, we will perform the integration by employing the partial wave expansion
\begin{align}
e^{k\cdot \hX}&=\sum_{l=0}^{\infty}\frac{d+2l}{d}|k|^{l}s_{l}(|k|)C_{l}^{(h)}\left(\frac{k\cdot \hX}{|k|}\right)\,,\label{eq:eikx-expansion-formula}\\
s_{l}(x)&=\sum_{j=l}^{\infty}\frac{\Ga\left(h+1\right)}{\Ga\left(h+1+j\right)}\frac{x^{2(j-l)}}{2^{2j-l}(j-l)!}\,.
\label{eq:sl-def}
\end{align}
From the partial wave expansion (\ref{eq:eikx-expansion-formula}), any function $f(\hX)$ can be formally expanded as
\begin{align}
f(\hX)&=\left(e^{\hX\cdot \partial_{y}}f(y)\right)\bigl\lvert_{y=0}
=\sum_{l=0}^{\infty}\frac{d+2l}{d}|\partial_{y}|^{l}s_{l}(|\partial_{y}|)C_{l}^{(h)}\left(\frac{\partial_{y}\cdot \hX}{|\partial_{y}|}\right)f(y)\Bigl\lvert_{y=0}\,,
\end{align}
where $|\partial_{y}|^2\equiv\partial_{y} \cdot \partial_{y}$\,.
By using this expansion and the integration formula
\begin{align}
\int_{S^{d+1}}\dd^{d+2}{\bf \hX}\,C_{l'}^{(h)}\left(\frac{k\cdot \hX}{|k|}\right)K_{l,0}^{\sfs}(\hX;T)
=\frac{d}{d+2l}\text{Vol}(S^{d+1})|k|^{-l}\,K_{l,0}^{\sfs}(k;T)\delta_{ll'}\,,
\end{align}
we obtain
\begin{align}
\int_{S^{d+1}}\dd^{d+2}{\bf \hX}\,f(\hX)K_{l,0}^{\sfs}(\hX;T)
=\text{Vol}(S^{d+1})\left(s_{l}\left(|\partial_{y}|\right)K_{l,0}^{\sfs}(\partial_{y};T)f(y)\right)\Bigl\lvert_{y=0}\,.
\label{eq:int-formula}
\end{align}
The three-point function (\ref{eq:three-point-KKK}) corresponds to the case $f(\hX)=K_{l_2,0}^{\sfs}(\hX;T_2)K_{l_3,0}^{\sfs}(\hX;T_3)$\,.

\medskip

Now let us use the formula (\ref{eq:int-formula}) to evaluate (\ref{eq:three-point-KKK}).
Note that the term in the expansion of $s_{l}(|\partial_{y}|)$ gives non-vanishing contribution only if $j=L_{123}=\frac{l_1+l_2+l_3}{2}$\,.
Therefore, $\hA_{l_1l_2l_3}(T_i)$ can be rewritten as
\begin{align}
\hA_{l_1l_2l_3}(T_i)&=\text{Vol}(S^{d+1})\left(s_{l_1}\left(\partial_{y}^2\right) K_{l_1,0}^{\sfs}(\partial_y;T_1)\left[K_{l_2,0}^{\sfs}(y;T_2)K_{l_3,0}^{\sfs}(y;T_3)\right]\right)\Bigl\lvert_{y=0}\no\\
&=\frac{\text{Vol}(S^{d+1})2^{l_1}\Ga\left(h+1\right)}{\Ga\left(h+1+L_{123}\right)}
\frac{1}{2^{2L_{123}}\alpha_{231}!}\no\\
&\qquad\qquad \times\left((\partial_{y}^2)^{\alpha_{231}}K_{l_1,0}^{\sfs}(\partial_y;T_1)[K_{l_2,0}^{\sfs}(y;T_2)K_{l_3,0}^{\sfs}(y;T_3)]\right)\bigl\lvert_{y=0}\,.
\end{align}
Note that this expression does not vanish when $\alpha_{231}$ is zero or positive integers.
Since $\partial^2_yK_{l_2,0}^{\sfs}(y;T_2)=\partial^2_yK_{l_3,0}^{\sfs}(y;T_3)=0$ for $y\in\mathbb{R}^{d+2}$\,, we obtain
\begin{align}
\hA_{l_1l_2l_3}(T_i)
&=\hC_{l_1,0}\hC_{l_2,0}\hC_{l_3,0}\frac{\text{Vol}(S^{d+1})\Ga\left(h+1\right)}{\Ga\left(h+1+L_{123}\right)}
\frac{l_2!l_3!T_{23}^{\alpha_{231}}}{2^{L_{123}}\alpha_{231}!\alpha_{312}!\alpha_{123}!}\no\\
&\qquad\times\left((T_1\cdot \partial_{y})^{l_1}[(T_2\cdot y)^{\alpha_{123}}(T_3\cdot y)^{\alpha_{312}}]\right)\bigl\lvert_{y=0}\no\\
&=\hC_{l_1,0}\hC_{l_2,0}\hC_{l_3,0}\frac{(-1)^{L_{123}}\text{Vol}(S^{d+1})\Ga\left(h+1\right)}{\Ga\left(h+1+L_{123}\right)}
\frac{l_1!l_2!l_3!}{\alpha_{231}!\alpha_{312}!\alpha_{123}!}
T_{23}^{\alpha_{231}}T_{31}^{\alpha_{312}}T_{12}^{\alpha_{123}}\,.
\end{align}
Since this formula is symmetric under the permutations of $l_{i}$\,, when $l_i \in \mathbb{Z}_{\geq0}$, $\hA_{l_1l_2l_3}(T_i)$ does not vanish only if $\alpha_{231}\,, \alpha_{312}\,, \alpha_{123}\in \mathbb{Z}_{\geq0}$\,.
The above proof is a higher dimensional generalization of the proof given in \cite{hyperspherical2}.

\subsection{Spinning case}
\paragraph{}
Next, let us see the tensor spherical harmonic function $K_{l,s}^{\sfs}(\hX,U;T,R)$ given in (\ref{def:tsh}),
\begin{align}
\begin{split}
  K_{l,s}^{\sfs}(\hX,U;T,R)&=\hC_{l,s}(\hX\cdot \hsfC\cdot U)^{s}(\hX\cdot T)^{l-s}\,,\\
 \hsfC_{\hA\hB}&=T_{\hA}R_{\hB}-R_{\hA}T_{\hB}\,.
 \end{split}
\end{align}
We can show that the polynomial (\ref{def:tsh}) satisfies the harmonic equation:
\begin{align}
\hnabla^2K_{l,s}^{\sfs}(\hX,U;T,R)=-(l(l+d)-s)K_{l,s}^{\sfs}(\hX,U;T,R)\,,
\label{eq:tsh-eign}
\end{align}
and the homogeneity and the transversality condition
\begin{align}
(\hX\cdot \partial_{\hX}-l)K_{l,s}^{\sfs}(\hX,U;T,R)=0\,,\qquad
\hX\cdot \partial_U K_{l,s}^{\sfs}(\hX,U;T,R)=0\,.
\label{eq:TSH-con}
\end{align}
The eigenvalue (\ref{eq:tsh-eign}) of the Laplacian precisely matches with the one of the rank-$s$ tensor spherical harmonic function on $S^{d+1}$ \cite{hyperspherical4}.
The harmonic function (\ref{def:tsh}) also obeys the orthogonality relation (\ref{eq:ortho-tsh}).

\medskip

The equation (\ref{eq:tsh-eign}) implies that the harmonic function is an eigenfunction of the quadratic Casimir operator constructed from the $SO(d+2)$ generators 
\begin{align}
\hat{\cL}_{\hA\hB}\equiv \hat{X}_{\hA}\frac{\partial}{\partial \hX_{\hB}}-\hat{X}_{\hB}\frac{\partial}{\partial \hX_{\hA}}+\hat{\Sigma}_{\hA\hB}\,.
\end{align}
In fact, the action of the quadratic Casimir operator for $SO(d+2)$ can be evaluated as
\begin{align}
\frac{1}{2}\hat{\cL}_{\hA\hB}\hat{\cL}^{\hA\hB}K_{l,s}^{\sfs}(\hX,U;T,R)&=\left(-\hat{\nabla}^2-s+\frac{1}{2}\Sigma_{\hA\hB}\Sigma^{\hA\hB}\right)K_{l,s}^{\sfs}(\hX,U;T,R)\no\\
&=\left(l(l+d)+s(s+d-2)\right)K_{l,s}^{\sfs}(\hX,U;T,R)\,,
\label{eq:TSH-eigen}
\end{align}
where we used the relation
\begin{align}
\frac{1}{2}\Sigma_{\hA\hB}\Sigma^{\hA\hB}K_{l,s}^{\sfs}(\hX,U;T,R)
=-s(s+d-2)K_{l,s}^{\sfs}(\hX,U;T,R)\,.
\end{align}
Note that when $d=4$\,, the eigenvalue (\ref{eq:TSH-eigen}) of the $SO(6)$ quadratic Casimir operator corresponds to the $SU(4)$ representation $[s,l-s,s]$\,.

\medskip

Finally, as in the AdS case \cite{Sleight:2016hyl}, we will give a formula which connects the tensor spherical harmonic function (\ref{def:tsh}) to the SSH (\ref{def:shf}).
Such formula is given by
\begin{align}
K_{l,s}^{\sfs}(\hX,U;T,R)&=\frac{\Gamma(l-s+1)\hC_{l,s}}{\Gamma(l+1)\hC_{l,0}}
\left(\hcD_{T}(U,R)\right)^s K_{l,0}^{\sfs}(\hX;T)\,,
\label{eq:spin-scalar-s}
\end{align}
where the differential operator $\hat{\cD}_{T}(U,R)$ is defined as
\begin{align}
\hcD_{T}(U,R)=-(U\cdot R)\left(R\cdot \frac{\partial}{\partial R}-T\cdot \frac{\partial}{\partial T}\right)
-(U\cdot T)\left(R\cdot \frac{\partial}{\partial T}\right)\,.
\label{cDT-def}
\end{align}
Note that $\hat{\cD}_{T}(U,R)(\hX\cdot \hat{\sfC}\cdot U)^k=0$ for any constant $k$\,.

\subsubsection*{Orthogonality relation}
\paragraph{}
Finally, let us show the orthogonality relation of the tensor spherical harmonic functions,
\begin{align}
\frac{1}{\left(\frac{d-1}{2}\right)^{s}}\int_{S^{d+1}}\dd^{d+2}{\bf \hX}\,K_{l,s}^{\sfs}(\hX,\hat{K};T_1,R_1)K_{l',s'}^{\sfs}(\hX,U;T_2,R_2)\propto \delta_{ll'}\delta_{ss'}\hat{\mathsf{H}}_{12}^{s}T_{12}^{l-s}\,,
\label{eq:ortho-tsh}
\end{align}
where $\hat{\mathsf{H}}_{12}$ is defined as
\begin{align}
\hat{\mathsf{H}}_{12}\equiv
\hsfC_{2}\cdot \hsfC_{1}^{T}\equiv \hsfC_{1,\hA\hB}\hsfC_{2}^{\hA\hB}
=2[(T_{1}\cdot T_{2})(R_{1}\cdot R_{2})-(R_1\cdot T_1)(T_1\cdot R_2)]\,.\label{def:H}
\end{align}
Note that the orthogonality relation (\ref{eq:ortho-tsh}) has the similar form with spinning two-point functions of CFT.

\medskip

In order to show (\ref{eq:ortho-tsh}), we first note that since $K_{l,s}^{\sfs}(\hX,\hat{K};T_1,R_1)$ satisfies the transversality condition (\ref{eq:TSH-con}), the differential operator $\hat{K}$ in the integral (\ref{eq:ortho-tsh}) effectively acts on $K_{l',s'}^{\sfs}(\hX,U;T_2,R_2)$ like $\frac{d-1}{2}\frac{\partial}{\partial U^{\hat{A}}}$\,.
By applying (\ref{eq:spin-scalar-s}) to the left-hand side of (\ref{eq:ortho-tsh}) and using (\ref{eq:orthogonal-S}), we obtain:
\begin{align}
&\text{RHS of }(\ref{eq:ortho-tsh})\no\\
&=Z_{l}\,\delta_{ll'}\delta_{ss'}
\frac{\Gamma(l-s+1)^2\hC_{l,s}^2}{s!\Gamma(l+1)^2}(\partial_{U}\cdot\partial_{\bar{U}})^s
\left(\hcD_{T_1}(\bar{U},R_1)\right)^s
\left(\hcD_{T_2}(U,R_2)\right)^{s}(T_1\cdot T_2)^{l}\no\\
&=Z_{l}\,\delta_{ll'}\delta_{ss'}
\frac{\Gamma(l-s+1)\hC_{l,s}^2}{s!\Gamma(l+1)}(\partial_{U}\cdot\partial_{\bar{U}})^s
\left(\hcD_{T_1}(\bar{U},R_1)\right)^s
\left((T_1\cdot \hsfC_2\cdot U)^{s}(T_1\cdot T_2)^{l-s}\right)\no\\
&=Z_{l}\,\delta_{ll'}\delta_{ss'}\frac{\Gamma(l-s+1)\hC_{l,s}^2}{s!\Gamma(l+1)}
\sum_{r=0}^{s}\frac{s!}{r!(s-r)!}
\frac{\Gamma(s+1)\Gamma(l-s+1)}{\Gamma(s-r+1)\Gamma(l+r+1)}\no\\
&\quad\times 
(\partial_{U}\cdot\partial_{\bar{U}})^s
[(\bar{U}\cdot \hsfC_1^{T}\cdot \hsfC_2\cdot U)^r(T_1\cdot \hsfC_2 \cdot U)^{s-r}(\bar{U}\cdot \hsfC_1^{T}\cdot T_2)^{s-r}]\,(T_{1}\cdot T_{2})^{l+r-2s}\no\\
&\propto \delta_{ll'}\delta_{ss'}\hat{\mathsf{H}}_{12}^{s}T_{12}^{l-s}
\,.
\label{eq:ortho-tsh1}
\end{align}
In the final equation, we used (\ref{def:H}) and
\begin{align}
&(T_1\cdot \hsfC_2\cdot \hsfC_1^{T})_{\hat{A}}
=\frac{1}{2}\hat{\mathsf{H}}_{12}T_{1,\hat{A}}\,,\\
&(T_1\cdot \hsfC_2)\cdot (\hsfC_1^{T}\cdot \hsfC_2)^{p}\cdot(\hsfC_1^{T}\cdot T_2) =
\frac{1}{2^{p}}\hat{\mathsf{H}}_{12}^{p}T_{1}\cdot T_{2}\,.
\end{align}

\section{Exton function}\label{sec:Extonfun}
\paragraph{}
In this appendix, we will summarize some properties of the so called  Exton function \cite{Exton} (see also \cite{Fortin:2019dnq}), which recently becomes increasingly relevant in constructing conformal blocks \cite{Chen:2019gka}.
This function is defined as a double power series
\begin{align}
G(\alpha,\beta,\gamma,\delta;x,y)=\sum_{m,n=0}^{\infty}\frac{(\delta-\alpha)_{m}(\delta-\beta)_{m}}{m!(\gamma)_{m}}
\frac{(\alpha)_{m+n}(\beta)_{m+n}}{n!(\delta)_{2m+n}}x^my^n\,.
\label{def:Exton2}
\end{align}
This double power series can be expressed as a linear combination of two Appell hypergeometric function $F_{4}$\,,
\begin{align}
&G(\alpha,\beta,\gamma,\delta;x,1-y)=\frac{\Gamma(\delta)\Gamma(\delta-\alpha-\beta)}{\Gamma(\delta-\alpha)\Gamma(\delta-\beta)}F_{4}(\alpha,\beta;\ga,\alpha+\beta+1-\delta;x,y)\no\\
&\qquad\qquad+\frac{\Gamma(\delta)\Gamma(\alpha+\beta-\delta)}{\Gamma(\alpha)\Gamma(\beta)}
y^{\delta-\alpha-\beta}F_{4}(\delta-\alpha,\delta-\beta;\ga,\delta-\alpha-\beta+1;x,y)\,,
\label{G-twoF4}
\end{align}
where Appell hypergeometric function $F_{4}$ is defined as
\begin{align}
F_4(\alpha,\beta;\ga,\ga';x,y)\equiv \sum_{m=0}^{\infty} \sum^{\infty}_{n=0}\frac{(\alpha)_{m+n}(\beta)_{m+n}}{m!n!(\ga)_m(\ga')_n}x^my^n\,,\qquad |x|^{\frac{1}{2}}+|y|^{\frac{1}{2}}<1\,.
\end{align}
This hypergeometric function satisfies the following partial differential equations:
\begin{align}
\mathbb{D}_1^{\alpha,\beta,\gamma}F_4(\alpha,\beta;\ga,\ga';x,y)=0\,,\qquad 
\mathbb{D}_2^{\alpha,\beta,\gamma'}F_4(\alpha,\beta;\ga,\ga';x,y)=0\,,\label{eq:Appell-eq}
\end{align}
where the differential operators $\mathbb{D}_1^{\alpha,\beta,\gamma}$ and  $\mathbb{D}_2^{\alpha,\beta,\gamma'}$ are defined by
\begin{align}
\mathbb{D}_1^{\alpha,\beta,\gamma}&=x(1-x)\frac{\partial^2}{\partial x^2}-y^2\frac{\partial^2}{\partial y^2}
-2 x\,y\, \frac{\partial^2}{\partial x\partial y}+(\gamma-(\alpha+\beta+1)\,y)\frac{\partial}{\partial y}-\alpha\beta\,, \label{def:D1-F4}\\
\mathbb{D}_2^{\alpha,\beta,\gamma'}&=y(1-y)\frac{\partial^2}{\partial y^2}-x^2\frac{\partial^2}{\partial x^2}
-2 x\,y\, \frac{\partial^2}{\partial x\partial y}+(\gamma'-(\alpha+\beta+1)\,x)\frac{\partial}{\partial x}-\alpha\beta
\label{def:D2-F4}\,.
\end{align}

\medskip

In the special parameters, the Exton function (\ref{def:Exton2}) has simple forms which are expressed in terms of the hypergeometric functions
\begin{align}
&G(\alpha,\beta,\gamma,\gamma;x,1-y)={}_2F_1(\alpha,\beta;\gamma;z){}_2F_1(\alpha,\beta;\gamma;w)\,,\label{eq:G-2d}\\
&G(\alpha,\beta,\gamma,\gamma+1;x,1-y)\no\\
&\qquad=\frac{z\,{}_2F_1(\alpha,\beta;\gamma+1;z){}_2F_1(\alpha-1,\beta-1;\gamma; w)-(z\leftrightarrow w)}{z-w}\no\\
&\qquad=\frac{z\,{}_2F_1(\alpha,\beta;\gamma+1;z){}_2F_1(\alpha-1,\beta-1;\gamma-1; w)-(z\leftrightarrow w)}{z-w}
\,,
\label{eq:G-4d}
\end{align}
where $z$ and $w$ are defined by
\begin{align}
x=z\,w\,,\qquad y=(1-z)(1-w)\,.
\end{align}
For derivations of (\ref{eq:G-2d}) and the first equality of (\ref{eq:G-4d}), see Appendix C in \cite{Dolan:2000uw}.
The second equality of (\ref{eq:G-4d}) can easily be shown by considering the difference between the first and second equations.

\subsection*{The Exton function and the quadratic Casimir equation}
\paragraph{}
As shown in subsection \ref{sec:R-symb}, the $R$-symmetry block (and the conformal block) are expressed by the Exton function.
For future reference, we will explicitly show that the expression (\ref{eq:sR-block-G-exp}) of the $R$-symmetry block with $s=0$ satisfies the quadratic Casimir equation,
\begin{align}
        \hat{\cD}_{d,\sigma,\tau}^{(\hat{a},\hat{b})}\hat{G}_{l,0}^{\hat{a},\hat{b}}(\sigma,\tau)=\frac{1}{2}l(l+d)\hat{G}_{l,0}^{\hat{a},\hat{b}}(\sigma,\tau)\,.
        \label{eq:Casimir-eq-SO}
\end{align}
Here, the differential operator $\hat{\cD}_{d,\sigma,\tau}^{(\hat{a},\hat{b})}$ for the $SO(d+2)$ part is defined by
\begin{align}
-\hat{\cD}_{d,\sigma,\tau}^{(\hat{a},\hat{b})}&=\frac{\hat{\cL}^2}{2}
-(\hat{a}+\hat{b})\left((1+\sigma-\tau)\left(\sigma\frac{\partial}{\partial \sigma}+\tau\frac{\partial}{\partial \tau}\right)
-(1-\sigma-\tau)\frac{\partial}{\partial \tau}\right)\no\\
&\qquad+\hat{a}\hat{b}(1+\sigma-\tau)\,,
\label{eq:Cas-def-xy}
\end{align}
where  $\hat{\cL}^2$ is the quadratic Casimir operator for $SO(d+2)$\,, and the action on any function $f(\sigma,\tau)$ is
\begin{align}
\frac{1}{2}\hat{\cL}^2f(\sigma,\tau)&=
-\left((1-\tau)^2-\sigma(1+\tau)\right)\frac{\partial}{\partial \tau}\tau\frac{\partial}{\partial \tau}f(\sigma,\tau)
-(1-\sigma+\tau)\sigma \frac{\partial}{\partial \sigma}\sigma\frac{\partial}{\partial \sigma}f(\sigma,\tau)\no\\
&\quad+2(1+\sigma-\tau)\sigma \tau \frac{\partial^2}{\partial \sigma \partial \tau}f(\sigma,\tau)
+d\,\sigma \frac{\partial}{\partial \sigma}f(\sigma,\tau)\,.
\end{align}

\medskip

In order to show (\ref{eq:Casimir-eq-SO}), we first use the formula (\ref{G-twoF4}) of the Exton function.
Then, the $R$-symmetry block (\ref{eq:sR-block-G-exp}) with $s=0$ becomes
\begin{align}
&\hat{G}_{l,0}^{\hat{a},\hat{b}}(\sigma,\tau)=\sigma^{-\frac{l}{2}}\,
\biggl[\frac{\Gamma(-l)\Gamma(\hat{a}+\hat{b})}{\Gamma(-\frac{l}{2}+\hat{a})\Gamma(-\frac{l}{2}+\hat{b})}
F_{4}\left(-\frac{l}{2}-\hat{a},-\frac{l}{2}-\hat{b};1-l-h,1-\hat{a}-\hat{b};\sigma,\tau \right)\no\\
&\qquad+\frac{\Gamma(-l)\Gamma(-\hat{a}-\hat{b})}{\Gamma(-\frac{l}{2}-\hat{a})\Gamma(-\frac{l}{2}-\hat{b})}
\tau^{\hat{a}+\hat{b}}F_{4}\left(-\frac{l}{2}+\hat{a},-\frac{l}{2}+\hat{b};1-l-h,1+\hat{a}+\hat{b};\sigma,\tau\right)\biggr]
\,.
\label{eq:sR-block-G-exp-F4}
\end{align}
In fact, applying the differential operator $\hat{\cD}_{d,\sigma,\tau}^{(\hat{a},\hat{b})}$ to (\ref{eq:sR-block-G-exp-F4}), we obtain
\begin{align}
\hat{\cD}_{d,\sigma,\tau}^{(\hat{a},\hat{b})}\hat{G}_{l,0}^{\hat{a},\hat{b}}(\sigma,\tau)
&=\frac{\sigma^{-\frac{l}{2}}}{\tilde{c}_{l,0}}
\left(2 \sigma\,\mathbb{D}_1^{ -\frac{l}{2}-\hat{a},-\frac{l}{2}-\hat{b};1-l-h  }+(1-\sigma-\tau) 
\mathbb{D}_2^{-\frac{l}{2}-\hat{a},-\frac{l}{2}-\hat{b};-l}\right)\no\\
&\qquad \qquad \times F_4\left(-\frac{l}{2}-\hat{a},-\frac{l}{2}-\hat{b};1-l-h,1-\hat{a}-\hat{b};\sigma,\tau \right)\no\\
&\quad +\frac{\sigma^{-\frac{l}{2}}\tau^{\hat{a}+\hat{b}}}{\tilde{c}_{l,0}}
\left(2 \sigma\,\mathbb{D}_1^{ -\frac{l}{2}+\hat{a},-\frac{l}{2}+\hat{b};1-l-h  }
+(1-\sigma-\tau) \mathbb{D}_2^{-\frac{l}{2}+\hat{a},-\frac{l}{2}+\hat{b};-l}\right)\no\\
&\qquad \qquad \times F_4\left(-\frac{l}{2}+\hat{a},-\frac{l}{2}+\hat{b};1-l-h,1+\hat{a}+\hat{b};\sigma,\tau\right)\no\\
&\quad +\frac{1}{2}l(l+d)\hat{G}_{l,0}^{\hat{a},\hat{b}}(\sigma,\tau)
=\frac{1}{2}l(l+d)\hat{G}_{l,0}^{\hat{a},\hat{b}}(\sigma,\tau)\,,
\end{align}
where $\mathbb{D}_1^{\alpha,\beta,\gamma}$ and $\mathbb{D}_2^{\alpha,\beta,\gamma'}$ are differential operators defined in (\ref{def:D1-F4}), (\ref{def:D2-F4})\,, and in the second equality we used the relations (\ref{eq:Appell-eq}) of $F_4$\,.

\section{Computations of the split representations of harmonic functions} \label{subsec:HfunctionS}
\paragraph{}
In this appendix, we will perform the integrals of our split representation (\ref{def:tharmonic-split}) of the harmonic function $\hat{\Omega}_{l,s}(\hX_{1},U;\hX_{2},U')$, and show that the resulting expression takes the form (\ref{eq:Omega-form}).
As a byproduct of the evaluation of (\ref{def:tharmonic-split}), we obtain explicit expressions (\ref{eq;scalar-split-result}), (\ref{eq:harmonic-spin1}) of the harmonic functions with $s=0\,, 1$\,.
Furthermore, we will present the coefficient $g_{l,s}^{(0)}(z)$ (\ref{eq:split-tensor-result}) of $\hat{\Omega}_{l,s}(\hX_{1},U;\hX_{2},U')$ which is proportional to $U_{12}^{s}$\,.

\medskip

Let us start with the integration (\ref{def:tharmonic-split})
\begin{align}
\hat{\Omega}_{l,s}(\hX_{1},U;\hX_{2},U')=\frac{1}{\left(h-1\right)_{s}s!}\int_{\mathbb{C}^{d+2}} D^{d}T\,K_{l,s}^{\sfs}(\hX_{1},U;T,\hat{D}_{R})K_{\tilde{l},s}^{\sfs}(\hX_{2},U';T,R)\,.
\end{align}
In order to evaluate this integral, we use the following formula:
\begin{align}
\frac{1}{s!\left(h-1\right)_s}(f\cdot \hat{D}_{R})^{s}(f'\cdot R)^{s}=\frac{s!}{2^s\left(h-1\right)_s}(f^2f'{}^2)^{\frac{s}{2}}C_{s}^{\left(h-1\right)}\left(\frac{f\cdot f'}{|f||f'|}\right)\,,
\end{align}
where two arbitrary vectors $f\,, f'$ are independent of the null vector $R$\,.
Then, we obtain
\begin{align}
\hat{\Omega}_{l,s}&=\frac{2^{2h+s}s!\hC_{l,s}\hC_{\tilde{l},s}}{\left(h-1\right)_{s}}\int_{\mathbb{C}^{d+2}} D^{d}T\,
\frac{(T\cdot U_1)^{s}(T\cdot U_2)^{s}
C_{s}^{\left(h-1\right)}(t) }{(2\hX_{1}\cdot T)^{-l+s}(2\hX_{2}\cdot T)^{-\tilde{l}+s}}\,,
\label{eq:tharmonic-int1}
\end{align}
where $t=\hat{X}_1\cdot \hat{X}_2+2^{-2}\,\bar{t}$ and $\bar{t}$ is defined as
\begin{align}
\bar{t}&=\frac{(2T\cdot \hX_{2})(2T\cdot\hX_{1})}{(T\cdot U_1)(T\cdot U_2)}U_{12}
-2\frac{(2T\cdot\hX_{1})(\hX_{2}\cdot U_1)}{(T\cdot U_1)}
-2\frac{(2T\cdot\hX_{2})(\hX_{1}\cdot U_2)}{(T\cdot U_2)}
\,.
\end{align}
This integral has a similar form as  (202) in \cite{Costa:2014kfa}.

\medskip

Next, let us use the finite series form of the Gegenbauer polynomial
\begin{align}
C_{s}^{(h-1)}(x)=\sum_{k=0}^{\lfloor s/2\rfloor}(-1)^k\frac{(h-1)_{s-k}}{k!(s-2k)!}(2x)^{s-2k}\,.
\end{align}
Then, the integral (\ref{eq:tharmonic-int1}) becomes
\begin{align}
\hat{\Omega}_{l,s}
&=\hC_{l,s}\hC_{\tilde{l},s}
\sum_{r=0}^{s}\frac{2^{2h+r}s!}{(s-r)!}\,W_{l,s,r}
\left(\frac{1}{\left(h-1\right)_{s}}\sum_{k=0}^{\lfloor r/2\rfloor}\frac{(-1)^{k}(h-1)_{s-k}
 }{k!(r-2k)!}(2z)^{r-2k}\right)\no\\
&=\hC_{l,s}\hC_{\tilde{l},s}
\sum_{r=0}^{s}\frac{2^{2h+r}s!}{(s-r)!}\,W_{l,s,r} \,(h-1+s)_{-r}\,C_{r}^{(h-1+s-r)}(z)\,,
\label{eq:tharmonic-int20}
\end{align}
where we defined the integral
\begin{align}
W_{l,s,r}\equiv\int_{\mathbb{C}^{d+2}} D^{d}T\frac{ (T\cdot U_1)^{s}(T\cdot U_2)^{s}\bar{t}^{s-r}}{(2\hX_1\cdot T)^{-l+s}(2\hX_{2}\cdot T)^{-\tilde{l}+s} }\,.
\end{align}
This integral can be rewritten as
\begin{align}
W_{l,s,r}&=\sum_{k_1+k_2+k_3=s-r}\frac{(s-r)!}{k_1!k_2!k_3!}(-2)^{k_2+k_3}U_{12}^{k_1}(U_1\cdot \hat{X}_2)^{k_2}( U_2\cdot \hat{X}_1)^{k_3}\no\\
&\quad\times \int_{\mathbb{C}^{d+2}} D^{d}T\frac{ (T\cdot U_1)^{s-k_1-k_2}(T\cdot U_2)^{s-k_1-k_3}}{(2\hX_1\cdot T)^{-l+s-k_1-k_2}(2\hX_{2}\cdot T)^{-\tilde{l}+s-k_1-k_3} }\no\\
&=\sum_{k_1+k_2+k_3=s-r}\frac{(s-r)!}{k_1!k_2!k_3!}\frac{(-1)^{k_2+k_3}U_{12}^{k_1}(U_1\cdot \hat{X}_2)^{k_2}(U_2\cdot\hat{X}_1)^{k_3}}{(l-s+k_1+k_2+1)_{k_3}(\tilde{l}-s+k_1+k_3+1)_{k_2}}\no\\
&\quad\times \left(U_1\cdot \frac{\partial}{\partial \hat{X}_1}\right)^{k_3}\left(U_2\cdot \frac{\partial}{\partial \hat{X}_2}\right)^{k_2}W_{l,r}\,,
\label{eq:Wlsr-Wlr}
\end{align}
where the integral $W_{l,r}$ is defined as
\begin{align}
W_{l,r}\equiv\int_{\mathbb{C}^{d+2}} D^{d}T\frac{ (T\cdot U_1)^{r}(T\cdot U_2)^{r}}{(2\hX_1\cdot T)^{-l+r}(2\hX_{2}\cdot T)^{-\tilde{l}+r} }\,.
\label{def:Wlr}
\end{align}
As we will show later, the integral (\ref{def:Wlr}) is given by (\ref{eq:Wlr-r>0}), (\ref{eq:Wlr-r=0}).
By substituting (\ref{eq:Wlsr-Wlr}) and (\ref{eq:Wlr-r=0}) into (\ref{eq:Wlsr-Wlr}), we can see that the general form of the harmonic function $\hat{\Omega}_{l,s}$ is
\begin{align}
    \hat{\Omega}_{l,s}(\hX_1,U_1;\hX_2,U_2)&=\sum_{r=0}^{s}U_{12}^{s-r}(U_1\cdot \hat{X}_2)^r(U_2\cdot \hat{X}_1)^r\,g_{l,s}^{(r)}(z)\,,
\end{align}
where the coefficients $g_{l,s}^{(r)}(z)$ are functions of $z$\,.
When $s=1$\,, the explicit expressions of $g_{l,1}^{(0)}(z)$ and $g_{l,1}^{(1)}(z)$ are presented in (\ref{eq:harmonic-spin1-g}) , and we can show that the resulting harmonic function $\hat{\Omega}_{l,1}$ satisfies the equation of motion (\ref{eq:EOM-spin1}).

\medskip
   
In general, the explicit expressions of the coefficients $g_{l,s}^{(r)}(z)$ have complicated forms as expected from (\ref{eq:tharmonic-int20}), (\ref{eq:Wlsr-Wlr}), (\ref{eq:Wlr-r>0}) and (\ref{eq:Wlr-r=0}).
Therefore, for simplicity, let us focus on the leading coefficient $ g_{l,s}^{(0)}(z)$ that corresponds to take polarization vectors $U_{1}\,, U_2$ such that 
$\hX_{1}\cdot U_2=0=\hX_{2}\cdot U_1$\,.
This choice is equivalent to set $k_1=s-r$ and $k_2=k_3=0$ in (\ref{eq:Wlsr-Wlr}). By using (\ref{eq:tharmonic-int20}), (\ref{eq:Wlsr-Wlr}) and (\ref{eq:Wlr-r=0}), we can see that the coefficient $g_{l,s}^{(0)}(z)$ has the form
\begin{align}
g_{l,s}^{(0)}(z)
&=\text{Vol}(S^{d})\hC_{l,s}\hC_{\tilde{l},s}
\sum_{r=0}^{s}\frac{(-1)^r s!r!\Gamma(2h)(h)_{r}}
{(s-r)!\Gamma(2(h+r))}\frac{C_{l-r}^{(h+r)}(z)}{C_{l-r}^{(h+r)}(1)}\no\\
&\qquad\times 
\,(h-1+s)_{-r}\,C_{r}^{(h-1+s-r)}(z)\,.
\label{eq:tharmonic-int4}
\end{align}

\subsection*{A proof of (\ref{eq:Wlr-r>0}) and (\ref{eq:Wlr-r=0})}

Now, let us perform the integration in (\ref{def:Wlr}).
For this purpose, we introduce a Feynman parameter $\alpha$, 
\begin{align}
W_{l,r}=\frac{\Gamma(2h+2r)}{\Gamma(-l+r)\Gamma(-\tilde{l}+r)}\int_0^{\infty} \frac{\dd \alpha}{\alpha}\alpha^{-\tilde{l}+r}\int_{\mathbb{C}^{d+2}} D^{d}T\frac{ (T\cdot U_1)^{r}(T\cdot U_2)^{r}}{\left(2T\cdot Q\right)^{2h+2r}}\,,
\label{eq:tharmonic-int2}
\end{align}
where we defined $Q^{\hA}\equiv \hX^{\hA}_1+\alpha\, \hX_{2}{}^{\hA}$\,.
The integral over the null vector $T^{\hA}$ in (\ref{eq:tharmonic-int2}) can be performed by using the formula
\begin{align}
&\int_{\mathbb{C}^{d+2}} D^{d}T\,\frac{(T\cdot U_1)^r(T\cdot U_2)^r}{(2T\cdot Q)^{2h+2r}}
=
\begin{cases}
\frac{\text{Vol}(S^{d})}{2^{2h}(Q^2)^{h}}\qquad (r=0)\\
\frac{\text{Vol}(S^{d})\Gamma(2h)(h)_rr!}{2^{2h+r}\Gamma(2(h+r))}\left[\frac{(-1)^rU_{12}^r}{(Q^2)^{h+r}}+\frac{2^r(h+r)_r\alpha^r( U_1\cdot \hat{X}_2)^{r}( U_2\cdot \hat{X}_1)^r}{r!(Q^2)^{h+2r}}\right]\qquad (r\geq1)
\end{cases}
\,,
\label{eq:int-formula-null}
\end{align}
where $\text{Vol}(S^d)=\frac{2\pi^{h+\frac{1}{2}}}{\Gamma\left(h+\frac{1}{2}\right)}$.
The above integration formula is derived from 
\begin{align}
\int_{\mathbb{C}^{d+2}} D^{d}T\,\frac{T^{\hA_1}\cdots T^{\hA_r}}{(2T\cdot Q)^{2h+2r}}=
\frac{\pi^{h}(2h+2r)_{-h}Q^{\hA_1}\dots Q^{\hA_{2r}}}{(Q^2)^{h+2r}}
-\text{traces}\,.
\label{eq:Dt-int-formula}
\end{align}
Then, $W_{l,r}$ with $r\geq1$ is expressed as
\begin{align}
W_{l,r}&=\frac{\text{Vol}(S^{d})\Gamma(2h)(h)_rr!}{2^{2h+r}\Gamma(-l+r)\Gamma(-\tilde{l}+r)}\biggl[
(-1)^rU_{12}^r\int_0^{\infty} \frac{\dd \alpha}{\alpha}\frac{\alpha^{-\tilde{l}+r}}{(Q^2)^{h+r}}
\no\\ &\quad
+\frac{2^r(h+r)_r}{r!}( U_1\cdot \hat{X}_2)^{r}( U_2\cdot \hat{X}_1)^r \int_0^{\infty} \frac{\dd \alpha}{\alpha}\frac{\alpha^{-\tilde{l}+2r}}{(Q^2)^{h+2r}}\biggr]
\no\\
&=(-1)^r\text{Vol}(S^d)\frac{\Gamma(2h)(h)_{r}r!}{2^{2h+r}\Gamma(2(h+r)) }
\frac{C^{(h+r)}_{l-r}(z)}{C^{(h+r)}_{l-r}(1)}U_{12}^r\no\\
&\quad+\text{Vol}(S^d)\frac{\Gamma(2h)(h)_{r}(h+r)_r(-l+r)_r(-\tilde{l}+r)_r}{2^{2h}\Gamma(2(h+2r)) }
\frac{C^{(h+2r)}_{l-2r}(z)}{C^{(h+2r)}_{l-2r}(1)}( U_1\cdot \hat{X}_2)^{r}( U_2\cdot \hat{X}_1)^r\,,
\label{eq:Wlr-r>0}
\end{align}
where $C^{(\alpha)}_{n}(1)=\frac{(2\alpha)_{n}}{n!}$\,, and we used the integration formula 
\begin{align}
&\int_0^{\infty} \frac{\dd \alpha}{\alpha}\frac{\alpha^{b-c}}{(1+2x\alpha+\alpha^2)^{b}}=
\frac{\Gamma(b+c)\Gamma(b-c)}{\Gamma(2b)}{}_2F_1\left(b+c,b-c;b+\frac{1}{2};\frac{1-x}{2}\right)\,.
\label{eq:Int-for-2}
\end{align}
When $r=0$\,, $W_{l,0}$ is given by
\begin{align}
W_{l,0}
&=\text{Vol}(S^d)\frac{1}{2^{2h}}
\frac{C^{(h)}_{l}(z)}{C^{(h)}_{l}(1)}\,.
\label{eq:Wlr-r=0}
\end{align}

\subsubsection*{A proof of (\ref{eq:Dt-int-formula})}

Finally, we will show the formula (\ref{eq:Dt-int-formula}).
First of all, let us note that (\ref{eq:Dt-int-formula}) can be expressed in terms of the most simplest case 
\begin{align}
I(Q)\equiv\int_{\mathbb{C}^{d+2}} D^{d}T\,\frac{1}{(2T\cdot Q)^{2h}}=
\frac{\pi^{h}(2h)_{-h}}{(Q^2)^{h}}\,.
\label{eq:scalar-int-I}
\end{align}
Indeed, by differentiating $I(Q)$ with respect to $Q^{\hA}$\,, we obtain
\begin{align}
\int_{\mathbb{C}^{d+2}} D^{d}T\,\frac{T^{\hA_1}\cdots T^{\hA_r}}{(2T\cdot Q)^{2h+2r}}&=
\frac{\Gamma(2h)}{2^{r}\Gamma(2h+r)}\left(\prod_{i=1}^r \frac{\partial}{\partial Q_{\hA_i}}\right)I(Q)\no\\
&=
\frac{\pi^{h}(2h+2r)_{-h}Q^{\hA_1}\dots Q^{\hA_{2r}}}{(Q^2)^{h+2r}}
-\text{traces}\,.
\end{align}
Therefore, in the following discussion, we focus on a proof of the formula (\ref{eq:scalar-int-I}).

\medskip

To this end, we shall choose a gauge.
By using the scaling symmetry and the rotation symmetry $I(g\,Q)=I(Q)\,, g\in SO(d+2)$, the null vector $T^{\hA}$ can be taken as 
\begin{align}
T^{\hA}=e_{1}^{\hA}+i\,\xi^{\hA}\,,\qquad e_{1}=(1,0,\dots ,0)\,,\qquad e_{1}\cdot \xi=0\,,\qquad \xi^2=1\,.
\end{align}
Then the integral $I(Q)$ becomes
\begin{align}
I(Q)&=\int \dd^{d+1}\bar{\xi}\,\delta(\bar{\xi}^2-1)\,\frac{(2Q^1)^{-2h}}{\left(1+i\,\frac{\bar{\xi}\cdot \bar{Q}}{Q^1}\right)^{2h}}\no\\
&=(2Q^1)^{-2h}\sum^{\infty}_{r=0}
\begin{pmatrix}
2h+r-1\\
r
\end{pmatrix}
(-i)^r\frac{F_r(\bar{Q})}{(Q^1)^{r}}\,,\\
F_r(\bar{Q})&\equiv \int \dd^{d+1}\bar{\xi}\,\delta(\bar{\xi}^2-1)\,(\bar{\xi}\cdot \bar{Q})^{r}\,.
\end{align}
Here we introduced $(d+1)$-dimensional vectors $\bar{\xi}\,, \bar{Q}$ which satisfy $\xi=(0, \bar{\xi})\,, Q=(Q^1, \bar{Q})$\,, respectively.
The integral is computed as follows. 
From the $SO(d+1)$ symmetry $F_{r}(\bar{Q})$ $F_{r}(g\,\bar{Q})=F_r(\bar{Q})\,, g\in SO(d+1)$\,, this integral vanishes for odd $r$ and should take the following form:
\begin{align}
F_{2k}(\bar{Q})=F_{2k}^{(0)}(\bar{Q}^2)^{k}\,,\qquad k\in\mathbb{Z}\,,
\end{align}
where $F_{2k}^{(0)}$ is the overall constant. 
In order to determine this constant, let us evaluate the integral $F_{2k}(\bar{Q})$ for a particular choice of $\bar{Q}$ i.e. $\bar{Q}=\bar{e}_1=(1,\dots ,0)$\,.
If we define $t\equiv \bar{\xi}\cdot \bar{e}_1$\,, we obtain
\begin{align}
F_{2k}^{(0)}=F_{2k}(e_1)&=\text{Vol}(S^{d-1})\int_{-1}^{1}\dd t\,t^{2k}(1-t^2)^{\frac{d-2}{2}}\no\\
&=\text{Vol}(S^{d-1})\frac{\Gamma\left(k+\frac{1}{2}\right)\Gamma\left(h\right)}{\Gamma(h+k+\frac{1}{2})}\no\\
&=\text{Vol}(S^{d})\frac{\Gamma\left(k+\frac{1}{2}\right)\Gamma\left(h+\frac{1}{2}\right)}{\pi^{\frac{1}{2}}\Gamma(h+k+\frac{1}{2})}\,,
\end{align}
where in the final equality we used
\begin{align}
\text{Vol}(S^{d-1})=\text{Vol}(S^{d})\frac{\Gamma\left(h+\frac{1}{2}\right)}{\pi^{1/2}\Gamma(h)}\,.
\end{align}
Finally, by using the relation
\begin{align}
2^{-2h}
\begin{pmatrix}
2h+2k-1\\
2k
\end{pmatrix}
\frac{\Gamma\left(k+\frac{1}{2}\right)}{\Gamma(h+k+\frac{1}{2}) }
=
\begin{pmatrix}
h+k-1\\
k
\end{pmatrix}
\frac{\Gamma(h)}{2\Gamma(2h) }
\,,
\end{align}
we obtain
\begin{align}
I(Q)
&=\frac{\text{Vol}(S^{d})}{2\pi^{\frac{1}{2}}}(Q^1)^{-2h}\sum^{\infty}_{k=0}
\begin{pmatrix}
h+k-1\\
k
\end{pmatrix}
\left(-\frac{(\bar{Q}\cdot \bar{Q})}{(Q^1)^{2}}\right)^{k}=\frac{\pi^{h}(2h)_{-h}}{(Q^2)^{h}}\,.
\end{align}
This is the formula (\ref{eq:scalar-int-I}) we want.


\section{Explicit expressions of the $R$-symmetry block}\label{sec:Rblock}

In this appendix, we will give some explicit expressions of the $R$-symmetry blocks with $d=4$.
For future reference, we will compare our expression (\ref{eq:R-block-series}) with the known result \cite{Nirschl:2004pa,Dolan:2004mu}
\begin{align}
\hat{G}_{l,s}^{\hat{a},\hat{b}}(\sigma,\tau)
&=2N_{l+s+2}^{\hat{a},\hat{b}}N_{l-s}^{\hat{a},\hat{b}}\,(\alpha_1\alpha_2)^{\hat{a}}
P^{(-\hat{a}+\hat{b},-\hat{a}-\hat{b})}_{\frac{l+s}{2}+\hat{a},\frac{l-s}{2}+\hat{a}}(\hat{w}_1,\hat{w}_2)
\,,\label{eq:R-block-known}
\end{align}
where we have set $\hat{a}\leq \hat{b}$\,.
Our $R$-symmetry block (\ref{eq:R-block-series}) with $s=0$ is equivalent to (\ref{eq:R-block-known}) (see subsection \ref{sec:R-symb}), below we list it for few special values of $(l, \hat{a}, \hat{b})$ for illustrations.
When $s>0$, (\ref{eq:R-block-series}) is same as (\ref{eq:R-block-known}) up the overall constant.
As we will observe, the difference between (\ref{eq:R-block-series}) and (\ref{eq:R-block-known}) does not depend on $\hat{a}$\footnote{It would be interesting to understand this independence of $\hat{a}$ a bit better, we suspect it is related to our choice of overall normalization for $\hat{G}_{l, s}^{\hat{a}, \hat{b}}(\sigma, \tau)$ which is also independent of $\hat{a}$. }.

\subsection*{$s=0$ case}
$(l,\hat{a},\hat{b})=(1,\frac{1}{2},\frac{1}{2})$
\begin{align}
    \frac{(1-\alpha_1) (1-\alpha_2)}{(\alpha_1 \alpha_2)^{1/2}}\,.\label{eq:s=0-1}
\end{align}
$(l,\hat{a},\hat{b})=(2,0,0)$
\begin{align}
    \frac{2 \alpha_1 \alpha_2-3 (\alpha_1+\alpha_2)+6}{6 \alpha_1 \alpha_2}\,.\label{eq:s=0-2}
\end{align}
$(l,\hat{a},\hat{b})=(2,0,1)$
\begin{align}
    \frac{(\alpha_1-1) (\alpha_2-1)}{\alpha_1 \alpha_2}\,.
    \label{eq:s=0-3}
\end{align}
$(l,\hat{a},\hat{b})=(3,\frac{1}{2},\frac{1}{2})$
\begin{align}
    \frac{(\alpha_1-1) (\alpha_2-1) (\alpha_1 \alpha_2-2 (\alpha_1+\alpha_2)+6)}{6 (\alpha_1 \alpha_2)^{3/2}}\,.
    \label{eq:s=0-4}
\end{align}
$(l,\hat{a},\hat{b})=(3,\frac{1}{2},\frac{3}{2})$
\begin{align}
    \frac{(\alpha_1-1)^2 (\alpha_2-1)^2}{(\alpha_1 \alpha_2)^{3/2}}\,.
    \label{eq:s=0-5}
\end{align}
$(l,\hat{a},\hat{b})=(3,\frac{3}{2},\frac{3}{2})$
\begin{align}
    \frac{(\alpha_1-1)^3 (\alpha_2-1)^3}{(\alpha_1 \alpha_2)^{3/2}}\,.
    \label{eq:s=0-6}
\end{align}
$(l,\hat{a},\hat{b})=(4,0,0)$
\begin{align}
\frac{\alpha_1^2 (3 (\alpha_2-4) \alpha_2+10)-4 \alpha_1 (\alpha_2 (3 \alpha_2-16)+15)+10 ((\alpha_2-6) \alpha_2+6)}{60 (\alpha_1 \alpha_2)^2}\,.
\label{eq:s=0-7}
\end{align}
$(l,\hat{a},\hat{b})=(4,0,1)$
\begin{align}
\frac{(\alpha_1-1) (\alpha_2-1) (3 \alpha_1 \alpha_2-5 
(\alpha_1+\alpha_2)+10)}{10 (\alpha_1\alpha_2)^2}\,.
\label{eq:s=0-8}
\end{align}
$(l,\hat{a},\hat{b})=(4,1,1)$
\begin{align}
   \frac{(\alpha_1-1)^2 (\alpha_2-1)^2 (2 \alpha_1 \alpha_2-5(\alpha_1+\alpha_2)+20)}{20 (\alpha_1\alpha_2)^2}\,.
   \label{eq:s=0-9}
\end{align}
$(l,\hat{a},\hat{b})=(4,0,2)$
\begin{align}
   \frac{(\alpha_1-1)^2 (\alpha_2-1)^2}{(\alpha_1\alpha_2)^2}\,.
   \label{eq:s=0-10}
\end{align}
$(l,\hat{a},\hat{b})=(4,1,2)$
\begin{align}
    \frac{(\alpha_1-1)^3 (\alpha_2-1)^3}{(\alpha_1\alpha_2)^2}\,.
    \label{eq:s=0-11}
\end{align}
$(l,\hat{a},\hat{b})=(4,2,2)$
\begin{align}
    \frac{(\alpha_1-1)^4 (\alpha_2-1)^4}{(\alpha_1\alpha_2)^2}\,.
    \label{eq:s=0-12}
\end{align}

\subsection*{$s=1$ case}

$(l,\hat{a},\hat{b})=(1,0,0)$
\begin{align}
    \text{(\ref{eq:R-block-known})}&=\frac{-\alpha_1\alpha_2+\alpha_1+\alpha_2}{\alpha_1 \alpha_2}\,,\\
        (\ref{eq:R-block-series})&=-\frac{1}{2}\text{(\ref{eq:R-block-known})}\,.
        \label{eq:s=1-1}
\end{align}
$(l,\hat{a},\hat{b})=(2,\frac{1}{2},\frac{1}{2})$
\begin{align}
    \text{(\ref{eq:R-block-known})}&=-\frac{(\alpha_1-1) (\alpha_2-1)[]4 \alpha_1 \alpha_2-5 (\alpha_1+\alpha_2)]}{5 (\alpha_1 \alpha_2)^{3/2}}\,,\\
        (\ref{eq:R-block-series})&=-\frac{5}{12}\text{(\ref{eq:R-block-known})}\,.\label{eq:s=1-2}
\end{align}
$(l,\hat{a},\hat{b})=(3,0,0)$
\begin{align}
     \text{(\ref{eq:R-block-known})}&=-\frac{(\alpha_1-2) (\alpha_2-2) [\alpha_1 \alpha_2-(\alpha_1+\alpha_2)]}{4 (\alpha_1\alpha_2)^2}\,,\\
      (\ref{eq:R-block-series})&=-\frac{4}{9}\text{(\ref{eq:R-block-known})}\,.\label{eq:s=1-3}
\end{align}
$(l,\hat{a},\hat{b})=(3,0,1)$
\begin{align}
    \text{(\ref{eq:R-block-known})}&= -\frac{(\alpha_1-1) (\alpha_2-1) [\alpha_1 \alpha_2-(\alpha_1+\alpha_2)]}{(\alpha_1\alpha_2)^2}\,,\\
     (\ref{eq:R-block-series})&=-\frac{3}{8}\text{(\ref{eq:R-block-known})}\,.\label{eq:s=1-4}
\end{align}
$(l,\hat{a},\hat{b})=(3,1,1)$
\begin{align}
     \text{(\ref{eq:R-block-known})}&=-\frac{(\alpha_1-1)^2 (\alpha_2-1)^2 [2\alpha_1\alpha_2-3( \alpha_1+\alpha_2)]}{3 (\alpha_1\alpha_2)^2}\,,\\
      (\ref{eq:R-block-series})&=-\frac{3}{8}\text{(\ref{eq:R-block-known})}\,.\label{eq:s=1-5}
\end{align}
$(l,\hat{a},\hat{b})=(4,\frac{1}{2},\frac{1}{2})$
\begin{align}
    \text{(\ref{eq:R-block-known})}&=-\frac{1}{105 (\alpha_1 \alpha_2)^{5/2}}(1-\alpha_1) (1-\alpha_2) \bigl[35 \left(\alpha_1^2+\alpha_2^2\right)+12 (\alpha_1 \alpha_2)^2\no\\
    &\qquad\qquad-45\alpha_1 \alpha_2(\alpha_1+\alpha_2)+170 \alpha_1 \alpha_2-105 (\alpha_1+\alpha_2)\bigr]\,,\\
        (\ref{eq:R-block-series})&=-\frac{7}{16} \text{(\ref{eq:R-block-known})}\,.\label{eq:s=1-6}
\end{align}
$(l,\hat{a},\hat{b})=(4,\frac{1}{2},\frac{3}{2})$
\begin{align}
    \text{(\ref{eq:R-block-known})}&=-\frac{(1-\alpha_1)^2 (1-\alpha_2)^2 (6 \alpha_1 \alpha_2-7 (\alpha_1+\alpha_2))}{7 (\alpha_1 \alpha_2)^{5/2}}\,,\\
        (\ref{eq:R-block-series})&=-\frac{7}{20}\text{(\ref{eq:R-block-known})}\,.\label{eq:s=1-7}
\end{align}
$(l,\hat{a},\hat{b})=(4,\frac{3}{2},\frac{3}{2})$
\begin{align}
    \text{(\ref{eq:R-block-known})}&=-\frac{(1-\alpha_1)^3 (1-\alpha_2)^3 (4 \alpha_1 \alpha_2-7 (\alpha_1+\alpha_2))}{7 (\alpha_1 \alpha_2)^{5/2}}\,,\\
        (\ref{eq:R-block-series})&=-\frac{7}{20}\text{(\ref{eq:R-block-known})}\,.\label{eq:s=1-8}
\end{align}

\subsection*{$s=2$ case}

$(l,\hat{a},\hat{b})=(2,0,0)$
\begin{align}
    \text{(\ref{eq:R-block-known})}&= \frac{6 (\alpha_1 \alpha_2)^2-15 \alpha_1 \alpha_2 (\alpha_1+\alpha_2)+10 \left(\alpha_1^2+\alpha_2^2+ \alpha_1 \alpha_2\right)}{10 (\alpha_1 \alpha_2)^2}\,,\\
     (\ref{eq:R-block-series})&=\frac{5}{12}\text{(\ref{eq:R-block-known})}\,.\label{eq:s=2-1}
\end{align}
$(l,\hat{a},\hat{b})=(3,\frac{1}{2},\frac{1}{2})$
\begin{align}
    \text{(\ref{eq:R-block-known})}&=\frac{(1-\alpha_1) (1-\alpha_2) \left[3 (\alpha_1 \alpha_2)^2-9 \alpha_1 \alpha_2 (\alpha_1+\alpha_2)+7 \left(\alpha_1^2+\alpha_2^2+\alpha_1 \alpha_2\right)\right]}{7 (\alpha_1 \alpha_2)^{5/2}}\,,\\
        (\ref{eq:R-block-series})&=\frac{7}{24} \text{(\ref{eq:R-block-known})}\,.
        \label{eq:s=2-2}
\end{align}
$(l,\hat{a},\hat{b})=(4,0,0)$
\begin{align}
    \text{(\ref{eq:R-block-known})}&= \frac{1}{70 (\alpha_1 \alpha_2)^3}
    \bigl[9 (\alpha_1 \alpha_2)^3-35\left(\alpha_1^3+\alpha_2^3\right)-45 (\alpha_1+\alpha_2) (\alpha_1 \alpha_2)^2
   +70 \alpha_1 \alpha_2 \left(\alpha_1^2+\alpha_2^2\right)\no\\
    & \quad+160 (\alpha_1 \alpha_2)^2-175(\alpha_1 \alpha_2)(\alpha_1+\alpha_2)+70 \left(\alpha_1^2+\alpha_2^2+\alpha_1 \alpha_2\right) \bigr]\,,\\
     (\ref{eq:R-block-series})&=\frac{7}{24}\text{(\ref{eq:R-block-known})}\,.\label{eq:s=2-3}
\end{align}
$(l,\hat{a},\hat{b})=(4,0,1)$
\begin{align}
    \text{(\ref{eq:R-block-known})}&= \frac{(\alpha_1-1) (\alpha_2-1)
    \bigl[9 (\alpha_1 \alpha_2)^2-21 (\alpha_1 \alpha_2)(\alpha_1+\alpha_2)+14 \left(\alpha_1^2+\alpha_1 \alpha_2+\alpha_2^2\right) \bigr]}{14(\alpha_1\alpha_2)^3}\,,\\
     (\ref{eq:R-block-series})&=\frac{7}{30}\text{(\ref{eq:R-block-known})}\,.\label{eq:s=2-4}
\end{align}
$(l,\hat{a},\hat{b})=(4,1,1)$
\begin{align}
    \text{(\ref{eq:R-block-known})}&= \frac{(\alpha_1-1)^2 (\alpha_2-1)^2 \bigl[18 (\alpha_1 \alpha_2)^2-63 \alpha_1 \alpha_2 (\alpha_1+\alpha_2)+56 \left(\alpha_1^2+\alpha_1 \alpha_2+\alpha_2^2\right)\bigr]}{56 (\alpha_1 \alpha_2)^3}\,,\\
     (\ref{eq:R-block-series})&=\frac{7}{30}\text{(\ref{eq:R-block-known})}\,.\label{eq:s=2-5}
\end{align}




\bibliographystyle{JHEP}
\bibliography{harmonic_function_arxiv_v2}

\end{document}